\documentclass[12pt,preprint,epsf]{aastex}
\usepackage{graphicx}
\usepackage{epsfig} 
\usepackage{lscape} 

\slugcomment{Accepted for Astronomical Journal;}

\shorttitle{AO Imaging of Low-$z$ Quasar Absorbers}
\shortauthors{Chun et al.}

\begin{document}


\title{Adaptive Optics Imaging of Low-redshift Damped Lyman-alpha Quasar Absorbers}


\author{Mark R. Chun} 
\affil{Institute for Astronomy, University of Hawaii, Hilo, HI, 96720}
\author{Soheila Gharanfoli and Varsha P. Kulkarni}
\affil{Dept. of Physics and Astronomy, University of South Carolina, Columbia, SC 29208}
\and
\author{Marianne Takamiya}
\affil{Dept. of Physics and Astronomy, University of Hawaii, Hilo, HI 96720}
\email{}



\begin{abstract}

We have carried out a high angular resolution near-infrared imaging study of
the fields of 6 quasars with 7 strong  absorption line systems at $z < 0.5$,  
using the Hokupa'a adaptive optics system and the QUIRC near-infrared camera on
the Gemini-North telescope.   These absorption  systems include 4 classical
damped Lyman-alpha absorbers (DLAs), 2 sub-DLAs, and one Lyman-limit system.  
Images were obtained in the H or K$'$ filters with FWHM between $0\farcs2 -
0\farcs5$ with the goal of detecting  the absorbing galaxies and identifying
their morphologies.   Features are seen at projected separations of
$0\farcs5-16\farcs0$ from the quasars and all of the fields show features at
less than 2\arcsec separation. We find candidate absorbers in all of the seven
systems.  With  the assumption that some of these are associated with the
absorbers, the absorbers are low luminosity  $ \le 0.1 L_{H}^{*}$ or
$L_{K}^{*}$; we do  not find any large bright candidate absorbers in any
of our fields.  Some fields show compact features that are too faint for
quantitative morphology,  but could arise in dwarf galaxies.   


\end{abstract}





\section{INTRODUCTION}

Damped Lyman-alpha absorption lines in quasar spectra are believed to arise
from intervening galaxies and intergalactic matter at various cosmological
epochs.  The damped Lyman-alpha absorbers (hereafter DLAs) are  classically
defined  as quasar absorbers with $\log N$(H~I)$ > 20.3$ while absorbers with
$19.0 < \log  N$(H~I)$< 20.3$ are conventionally classified as sub-DLAs.  This
distinction is based on the observational constraints of an early spectroscopic
study \cite[]{Wolfe86}.  Since the Ly-$\alpha$  line shows damping wings even 
at $\log N$(H~I)$ \sim 18$ in this paper we will refer  to both the sub-DLA and
DLAs as DLA systems.   

At high redshifts the DLAs are believed to  contain a large fraction of the 
co-moving density of neutral hydrogen in galaxies and  possibly account for 
all of the stars visible today \cite[e.g.][]{Wolfe95,Peroux03}.  The evolution
of metallicities in these absorbers provide important  probes of the  chemical
enrichment and star formation history of the Universe
\cite[]{Khare04,Kulkarni05}.  Unfortunately the connection  between DLAs and
galaxies has not been clearly established.  To shed more light on this
connection, it is necessary to complement the  wealth of spectroscopic data on
these absorbers with information on their morphologies, luminosities, 
colors, and image structure from direct imaging. 

It has  proven hard to obtain this information for most DLAs.  The imaging of
high-$z$ DLAs has been very difficult and a large fraction of the attempts to 
detect the Ly-$\alpha$ emission from high-redshift  intervening ($z_{abs} <
z_{em}$) DLAs have produced either  non-detections or weak detections 
\cite[e.g.][]{Smith89,Hunstead90,Lowenthal95,Djorgovski96}.  Imaging studies of
low-$z$ DLAs have been more encouraging.  Although not always spectroscopically
confirmed to be the absorbers, galaxies in images  of low-redshift absorber
fields often show a variety of morphologies: spirals, irregulars, low  surface
brightness (LSB) galaxies (e.g., Steidel et al. 1994, 1995;  LeBrun et al.
1997; Bowen et al. 2001; Cohen 2001; Turnshek et al. 2001).  Most of these
previous searches had limiting flux sensitivity thresholds of $\sim 0.2L^{*}$
and thus could not rule out LSBs, while all of the near-IR searches lacked
adequate angular  resolution to rule out dwarf galaxies close to the line of
sight.  It is not clear which  of the several competing scenarios for DLAs are
valid: large, bright,  rotating proto-spirals (Wolfe  et al. 1986; Wolfe \&
Prochaska 1998; Prochaska \& Wolfe 1997b, 1998),  gas-rich dwarf galaxies (York
et al. 1986; Matteucci et al. 1997),  merging proto-galactic fragments with
cold dark matter (e.g., Haehnelt, Steinmetz, \& Rauch 1998; Maller et al.
2001), collapsing  halos with merging clouds   (e.g., McDonald \&
Miralda-Escud\'e 1999), or low-surface brightness  galaxies (Jimenez, Bowen, \&
Matteucci 1999). 


Here we present the first adaptive optics observations of low-redshift DLAs. We
have obtained near infrared images of seven absorbers at $ 0.1 < z < 1.3$
with the University of Hawaii Hokupa'a adaptive optics system and near-infrared
camera QUIRC on the Gemini-North telescope.  


We discuss the observations and data reductions in Section 2.  The  analysis of
the data are presented in Section 3 and the results from individual fields are
discussed in Section 4.  Finally, in Section 5 we characterize our sample of
low-redshift DLAs based on our measurements of the sizes, impact parameters,
and image structure.   Throughout this paper we assume $\Omega_{m} = 0.3$, 
$\Omega_{\Lambda} = 0.7$, and $h =0.73$. 

\section{OBSERVATIONS AND DATA REDUCTION}


\subsection{OBSERVATIONS}

Our sample consists of seven low-redshift absorption systems that have 
confirmed Lyman-alpha absorption features.  The redshift range  was constrained
so that hour long exposures would reach limiting magnitudes representative of
low surface brightness features at the redshift of the absorption systems.  The
most stringent observational constraint was set by the AO system's requirement
that the target field have a sufficiently bright wavefront reference source for
the adaptive optics system ($R \le 17$) to provide a useful correction.  In
total six fields with seven absorption systems were observed.  We present the
object field properties in Table 1 where we have preserved the nomenclature of
DLA and sub-DLA for reference.  

Between August 2000 and April 2001, we observed the fields around the six
quasars in the H or K$'$ filter (Wainscoat \& Cowie 1992) with the
now-decommissioned University of Hawaii 36-element curvature adaptive optics
(AO) system (Hokupa'a) (Graves et al. 2000) on the Gemini-North telescope.  In
all cases the quasar was used as the wavefront reference source for the AO
system.  The University of Hawaii near-infrared camera QUIRC (Hodapp et al.
1996), containing a $1024 \times 1024$ pixel HgCdTe detector, with a pixel
scale of $0\farcs020$/pixel, was used as the focal-plane imager.   All data
were taken in better than one-arcsecond seeing conditions with two fields
observed under photometric conditions.   Table 2 summarizes the observations.

Each field was observed as a dithered series of short exposures, each exposure
being 30-180 seconds long.   The sides of the dither  pattern were 5\arcsec~
resulting in a final mosaic image with an area approximately  $15 \arcsec
\times  15 \arcsec$ in the deepest exposure region.   The dither size of
5\arcsec~ was a compromise between having as large a deep-exposure field as
possible and having a separation large enough to generate a sky frame from the
on-source data.  After each cycle through a five-point dither pattern, we
offset the telescope by roughly $0\farcs1$ and repeated the dither pattern. 
This offset was used to ensure that groups of bad pixels could be removed in
the data reduction.   This combination of offsets and dithers was carried out
until we obtained the total exposure times. 

The final mosaic images have quasars with FWHM from 0\farcs2 to just under
0\farcs5.   This should be considered as a rough estimate of the image 
resolution since in general the quasar is not a point source but may contain a
bright host galaxy.

\subsection{IMAGE DATA REDUCTION}

Standard data reduction techniques for near infrared imaging were used to
produce the final images.  Individual frames were sky-subtracted and
flat-fielded.  Separate sky frames were constructed for each individual
exposure by averaging source-masked dithered frames taken within 10-20 minutes
of the individual exposure.  Flat-fields were constructed from dome flat images
with the lamps on and off.  We favor the use of the  dome flats over sky flats
because sky flats constructed from the object images do not account for the
emission from dust on the telescope and instrument surfaces.  This is
especially important in the K$'$ where the difference between the sky flats and
the dome flats were as much as several percent.  Bad pixels were identified as
hot pixels in short dark frames, as dead pixels in flat frames, and as pixels
with large standard deviations in either of the dark or flat sequences.

Individual frames were registered to the nearest pixel using the centroid
of the quasar and averaged excluding bad pixels and $\ge 3\sigma$ deviations in
the stack.  The integer pixel registration is more than sufficient for this
data since in all cases we have 7-20 pixels across the FWHM of the images.  No
correction was made for field distortions in the Gemini/Hokupaa/QUIRC images as
these are small \cite[]{Rigaut05}.

\subsection{SUBTRACTION OF THE QUASAR POINT SPREAD FUNCTION}

In order to properly study the area near the quasar, we need to remove the
contribution to the image of the quasar and its host galaxy. This problem
depends critically on our knowledge of both a point spread function (PSF) not
well described by an analytic function and on the intrinsic nature of the
quasar and its host galaxy.   This section describes the QSO-subtraction
techniques we used.   

First, during the observation we observed stars as PSF calibrations.  These PSF
calibration targets were chosen to have a similar right ascension,
declination,  and visual magnitude as the wavefront reference sources (e.g. the
quasars).   Observations of the PSF calibration fields were made interspersed
with those of the quasar fields to sample the  changes  in the intrinsic
atmospheric seeing.  In all cases,  the shape of the calibration PSF did not
match point sources in the quasar field.  This  occurred for several reasons. 
First, at the faint magnitudes of these guide stars the correction of the
adaptive optics  system depends strongly on the photon flux in the wavefront
sensor.  While we  attempted to select nearby stars with similar catalog
magnitudes, matching the wavefront  sensor counts was in practice quite
difficult and the wavefront sensor photon flux was typically matched to no
better than $10-20\%$.  Second, while we observed the PSF calibration fields
interleaved with the quasar field, the timescale for seeing changes can be as
short as a few minutes.  This is too short for the PSF calibration star 
observations to, in practice, be made under exactly the same atmospheric
conditions.  Third, the calibration stars were faint in the H/K$'$ bands and
required long exposure times to reach a SNR at large radii to make them useful
for PSF subtraction.  This was unrealistic given the overheads required.  

To circumvent these problems, we attempted three techniques to generate PSFs
directly from the data.   First, for quasar fields containing a stellar source
we have a PSF which was taken under identical atmospheric conditions.  However,
for all of the observed fields, the quasar is the brightest  source so the
signal-to-noise ratio at large radii in these stellar images did  not match
that of the quasar.  In addition, in the few fields with stellar sources,  the
sources happen to be only a few arcseconds from the quasar.  While this is well
within the corrected field of view of the adaptive optics system
\cite[]{Chun98},  extracting a clean PSF to a radius containing most of the
stellar flux proved difficult.   Knowledge of the PSF at these large radii is
important because the PSFs can have a significant fraction of energy outside
their core.  For  example, the average $50\% $ encircled energy diameter was
$0\farcs59 $ in the H band and $0\farcs28$  in   the K$'$ band.   

Given that the quasar is the brightest object in the observed fields, we tried
constructing a PSF from an azimuthally averaged image of the quasar itself.  
The azimuthally averaged profile is computed  directly from the final image
and, by construction, is not subject to differences in the guide star
brightness or to variations in the instrinsic seeing.    The resulting
PSF-subtracted image is similar to an unsharp mask but here the smoothing is
done azimuthally.   The technique assumes that the light near the quasar is
well represented by the azimuthal average.  Asymmetries in the PSF can arise 
from astrophysical sources (e.g. a host galaxy) as well as instrumental sources
(e.g. telescope  windshake).  In many cases the residual in the
azimuthally-averaged PSF subtracted image contained one pair of symmetric
positive flux lobes and another pair of symmetric negative flux lobes.  

Finally, we constructed models of the PSFs using a principle component analysis
or  Karhunen-Loeve (KL) decomposition.  The Karhunen-Loeve decomposition has
been previously used to quantify PSFs by \cite{Lupton01} and \cite{Lauer02}. 
\cite{Lauer02} suggested using  the KL decomposition to quantify the field
variations of the PSF of adaptive optics systems.  Here, we have applied this
technique to quantify the  modes in which the PSF varies with time.   The basic
idea is to construct a basis function that characterizes the  temporal
variations of the image of the quasar.   Any component in the final quasar
image well fit by this basis function is assumed to be due to the AO PSF
changing and is removed from the final quasar image.   The Karhunen-Loeve
decomposition provides the means to construct the basis function.  A $ 4\arcsec
\times 4\arcsec$ region centered on the quasar in each of the individual
reduced frames was binned $2\time2$ pixels, normalized, and centered on the
quasar.  From this set of images ($P_{i}$), we construct the basis function by
calculating the  eigenvectors and eigenvalues for the PSF covariance matrix: 

\begin{equation}
Covar_{i,j} =  < P_{i} P_{j} >  \label{covar}
\end{equation}

Each eigenvector represents a mode in the basis function and the eigenvalue 
represents the relative importance of each mode in the basis set.  Once the
basis function is determined, the quasar in the final image can be
reconstructed using the first few modes (typically $\sim 30$) with the largest
eigenvalues of the basis function.    These first few modes are by virtue of
the  Karhunen-Loeve expansion the modes with the most variance within the data
set.  In using this reconstructed model PSF, we can in principle erroneously
interpret some underlying parts of the true light from the quasar, host galaxy,
or absorber galaxy as a component of the PSF since some modes of the basis
function could be similar to these light distributions.  Since we have
constructed the basis set from the covariance matrix, the modes of the basis
set describe image structure that changes with time.  Faint diffuse objects are
less sensitive to changes in the image PSF than small unresolved objects.

For each quasar field we generated a KL basis set and then fit the image of the
quasar in the final mosaic image with the first 30 modes.  This model quasar 
was then subtracted from the field to look for objects close to the line of
sight of the quasar.  In  fields with known stellar components we also fit the
final stellar image with this same basis set as a measure of the residuals in
the PSF subtracted images.  

\begin{figure} \includegraphics[width=1.0\textwidth]{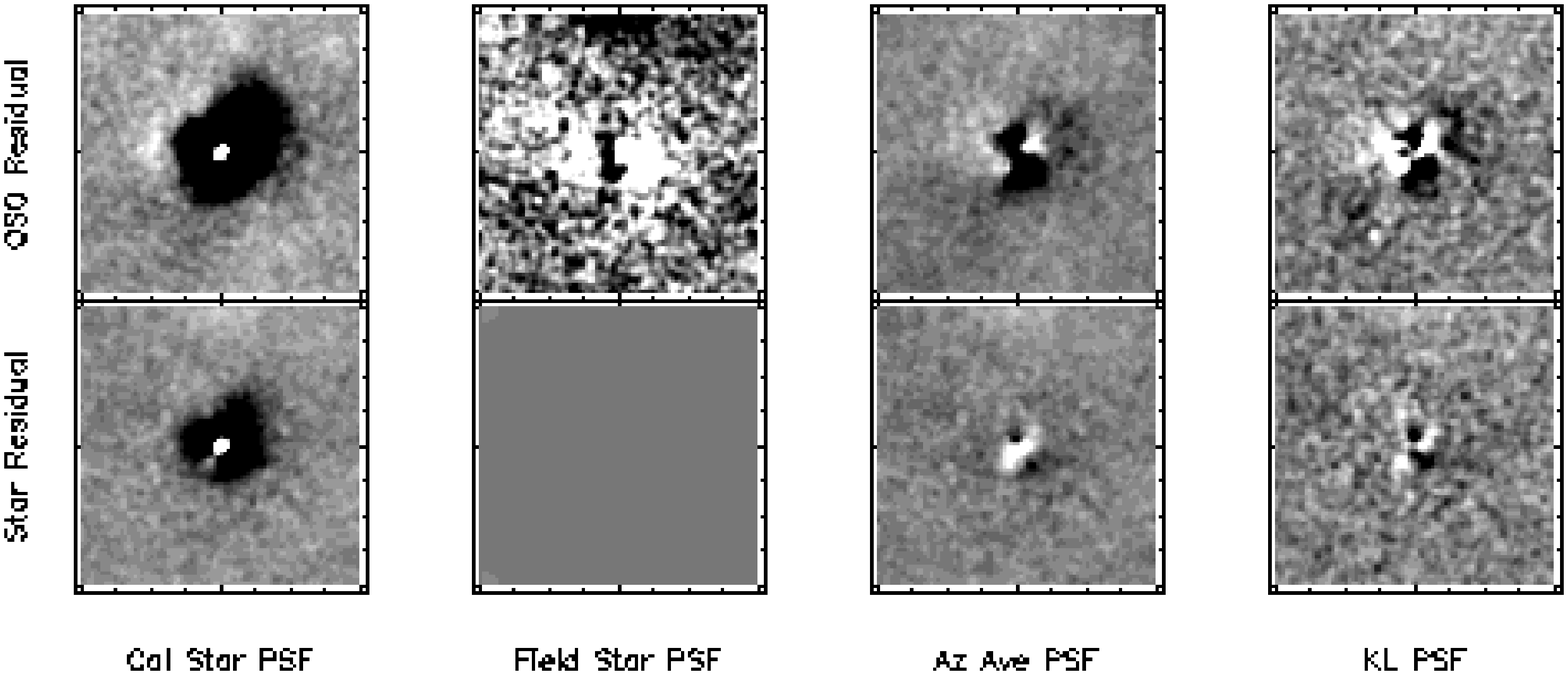}
\caption{ A comparison of the four techniques used to remove the contribution
of the quasar light to the image within the 4\arcsec $\times$ 4\arcsec~ region
centered on the quasar.   The axes are marked with 0.5\arcsec~ intervals. Each
column shows one of the techniques applied to the quasar (top) and a star
(bottom) in the field of Q0850+400.  The left most column shows the residual in
the quasar/star images when the PSF is constructed  from the PSF calibration
observations.  The second column shows the residuals when the PSF is 
constructed from the star within the field.  The field star residual image is
by definition zero.  The third column shows the residuals when the PSF is
constructed from the azimuthal average of the quasar image.  The last column
shows the residual when the PSF is modeled as a fit of the first 30 modes of
the Karhunen-Loeve (KL) basis calculated from the sequence of individual quasar
exposures in  the observation.  Residuals in the KL PSF stellar image 
subtraction extend to a radius of about 0\farcs5. In all cases, the PSF was
scaled and translated to minimize the  variance within the $4\arcsec \times
4\arcsec$  images.  All images are displayed with the maximum and minimum
intensities scaled to the sky $\pm0.1\%$ of the peak intensity in the
unsubtracted images.   The residual images have been  smoothed by a guassian
with FWHM=3 pixels.   \label{fig:PSFSubtraction}} \end{figure}

Figure \ref{fig:PSFSubtraction} shows a comparison of the techniques we used to
remove the  contribution of light from the quasar.  Each of the four columns in
the figure show a different technique applied to the quasar and to the
point-source just South of the quasar in the Q0850+440 field.  The cleanest 
quasar subtraction is using the KL PSF though this analysis is CPU intensive
and could only be applied within a couple of arcseconds from the quasar.
Residuals in the stellar image subtraction extend to a radius of about
0\farcs5. We take this radius as the minimum separation from the quasar line of
sight for an object to be detected with this technique.  As simple and as fast
as it is, the azimuthal averaged quasar image worked surprisingly well.  While
the intensities in these residual images were larger than those in the  KL
QSO-subtracted image by a factor of two, the extent of the residuals were
nearly identical.

\begin{figure}
\begin{tabular}{cc}
\includegraphics[width=0.5\textwidth]{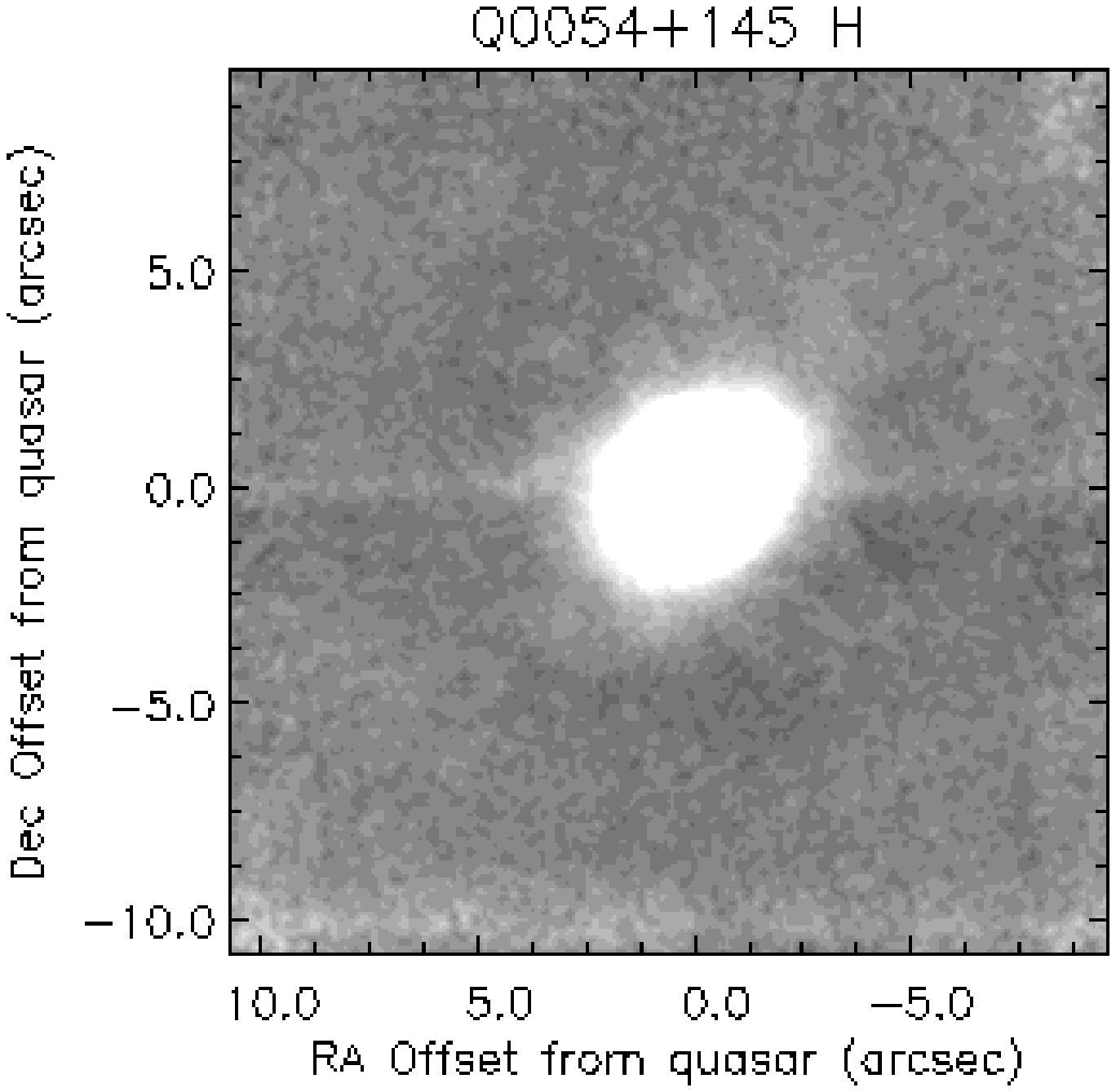}
\includegraphics[width=0.5\textwidth]{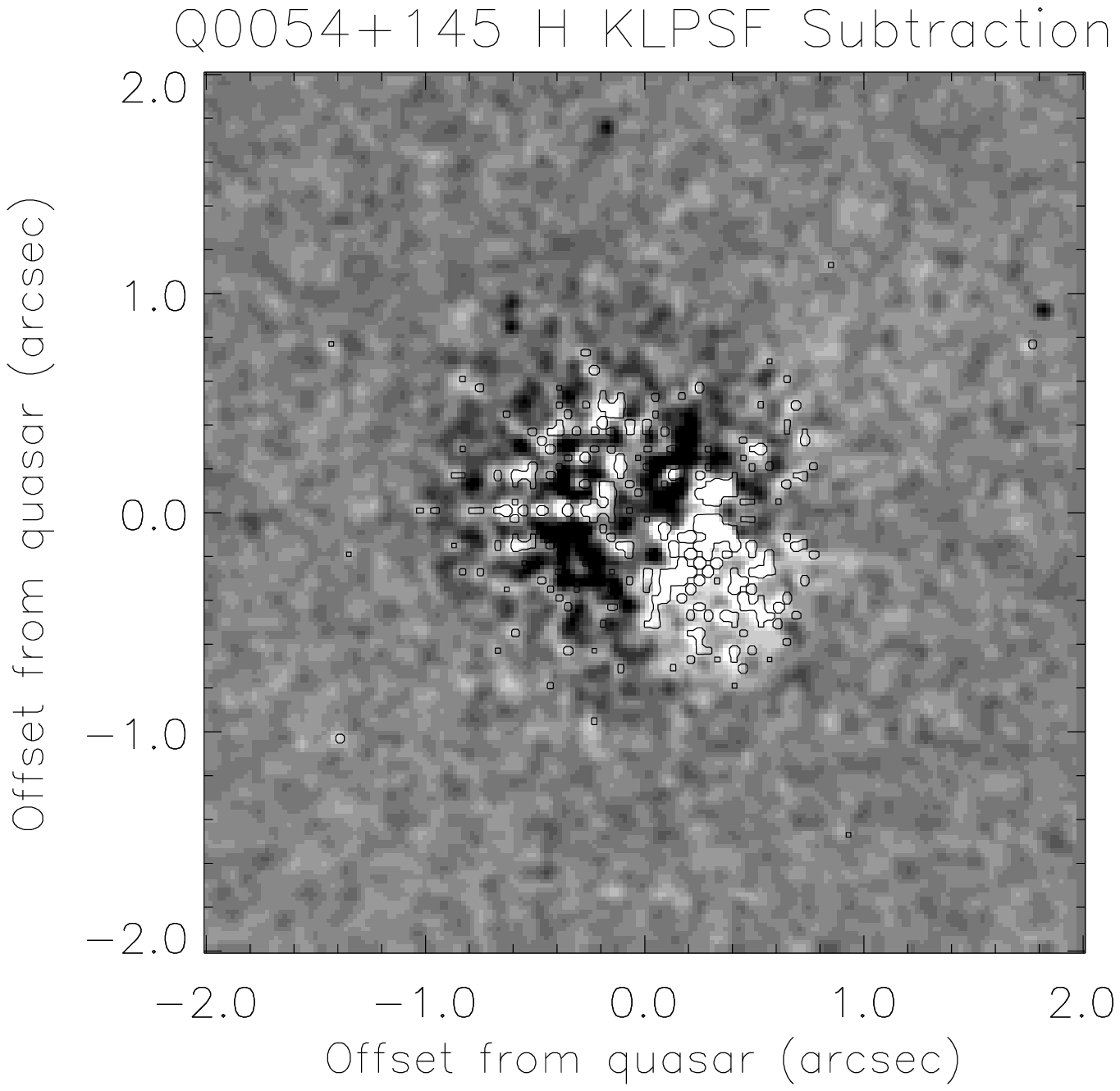}\\
\includegraphics[width=0.5\textwidth]{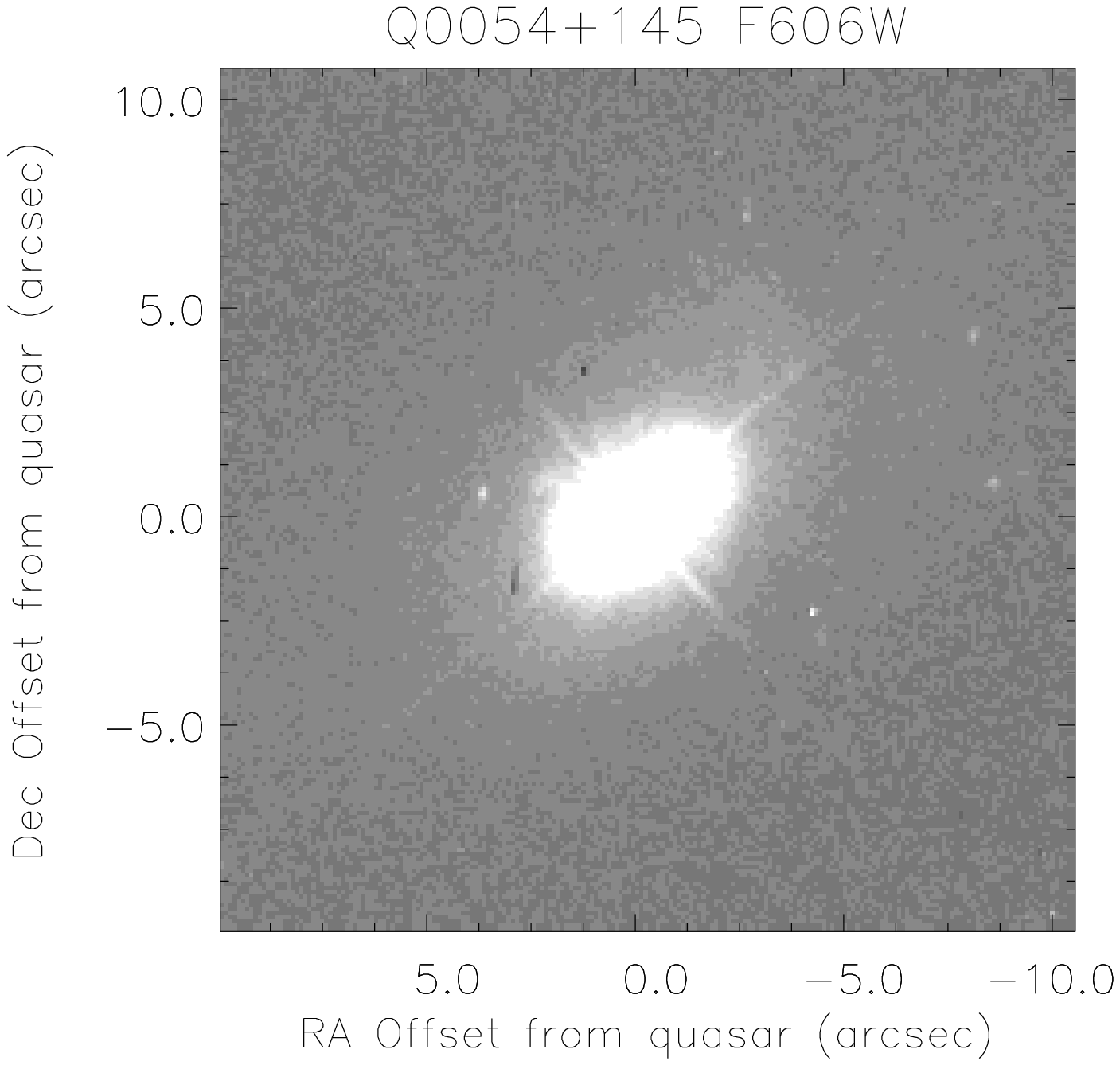}
\end{tabular}
\caption{Upper left panel shows our Gemini/Hokupaa H-band image
smoothed by a gaussian with FWHM=0\farcs2.  The upper-right
panel shows the quasar subtracted image of the central $4 \arcsec$ smoothed by
a 3-pixel gaussian.  Contours are overlaid on the image at 3$\sigma$ above the
sky in the unsmoothed image.  The bottom panel shows the HST/WPFC2 F606W image.  
All figures are shown in a linear intensity scale
with North up and East to the left.  The linear E-W feature running across
the upper-left image is an artifact caused by the detector quadrant boundary.
\label{fig:0054+145}}
\end{figure}

\begin{figure}
\begin{tabular}{cc}
\includegraphics[width=0.5\textwidth]{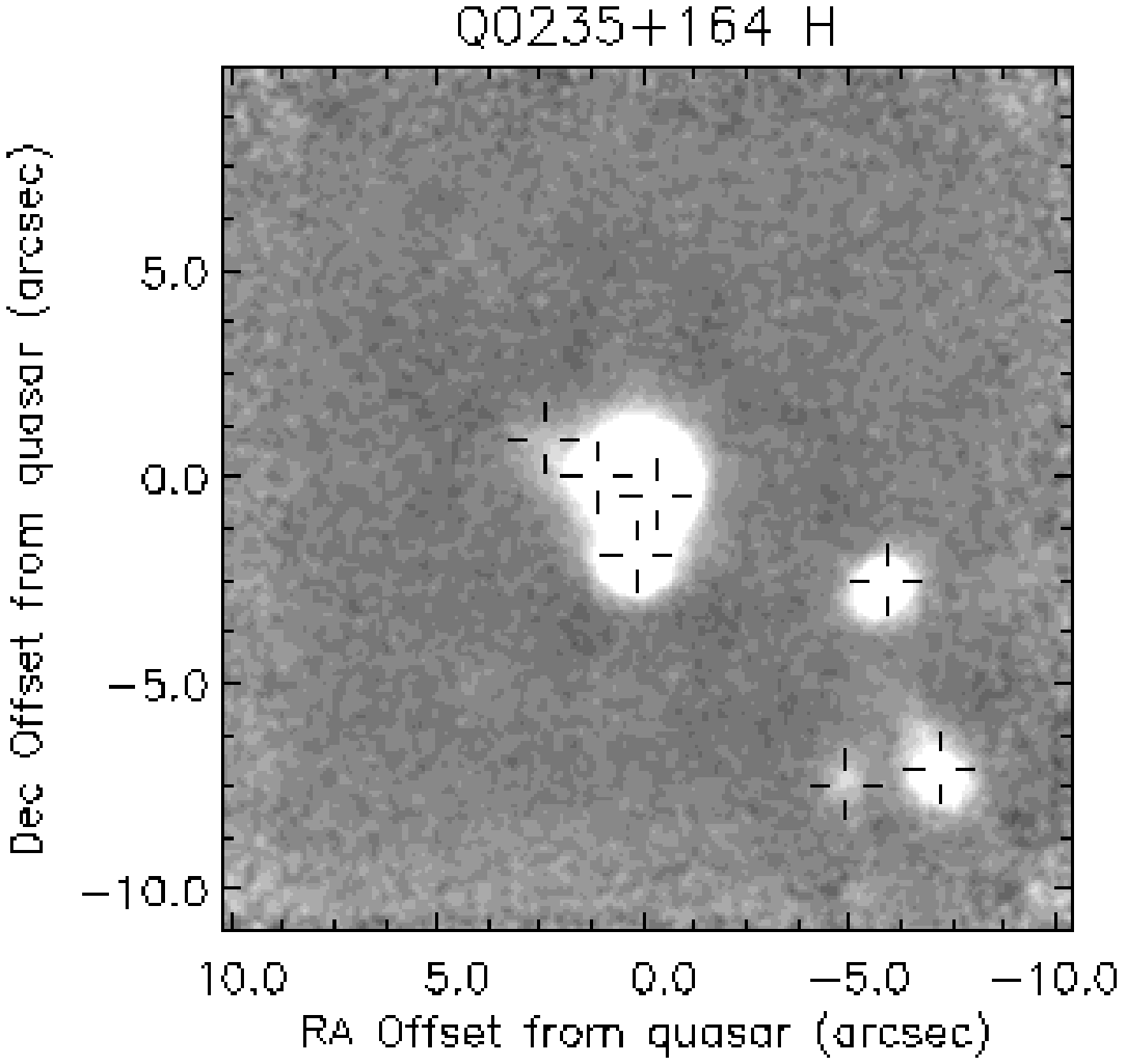}
\includegraphics[width=0.5\textwidth]{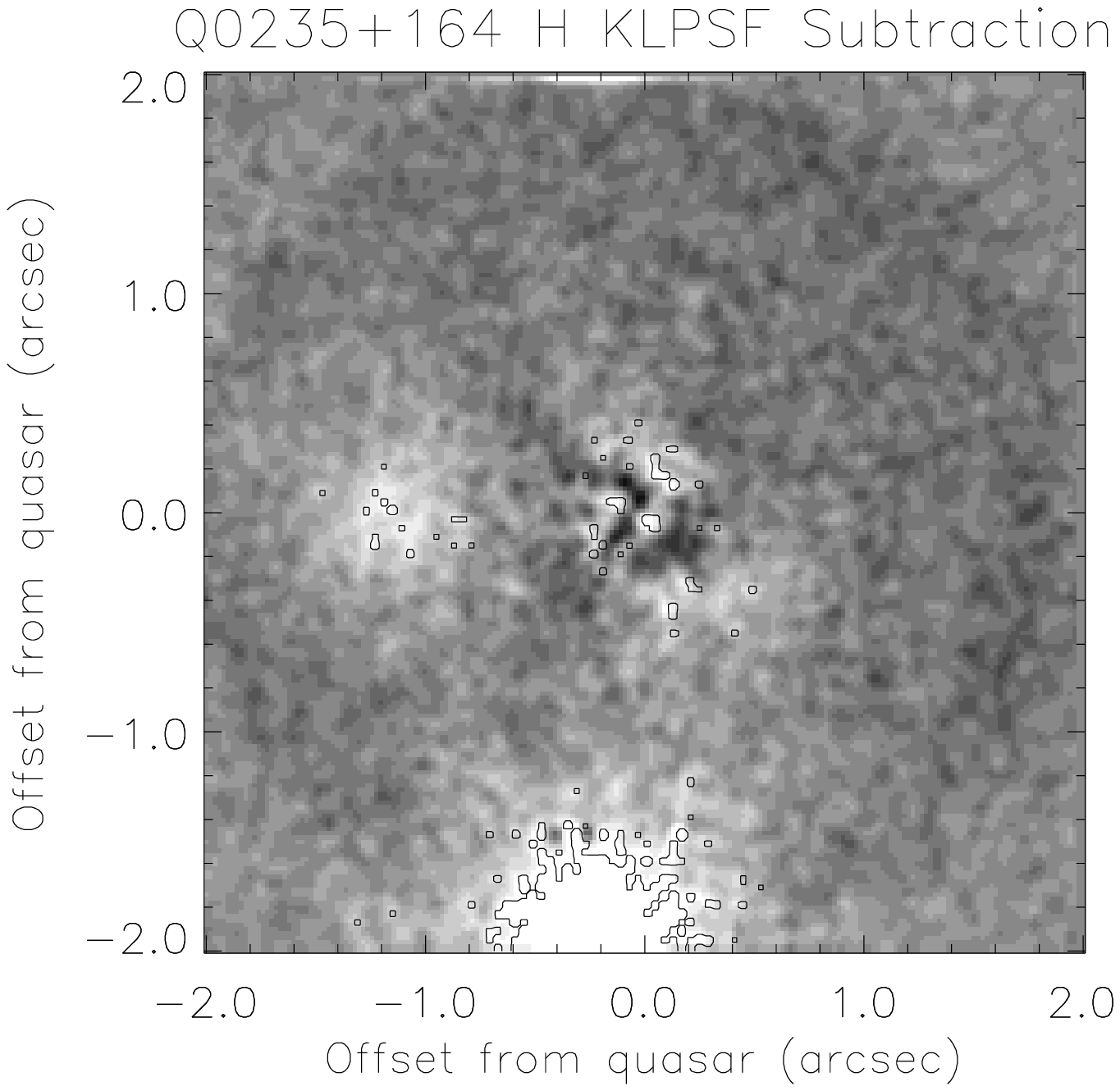}\\
\includegraphics[width=0.5\textwidth]{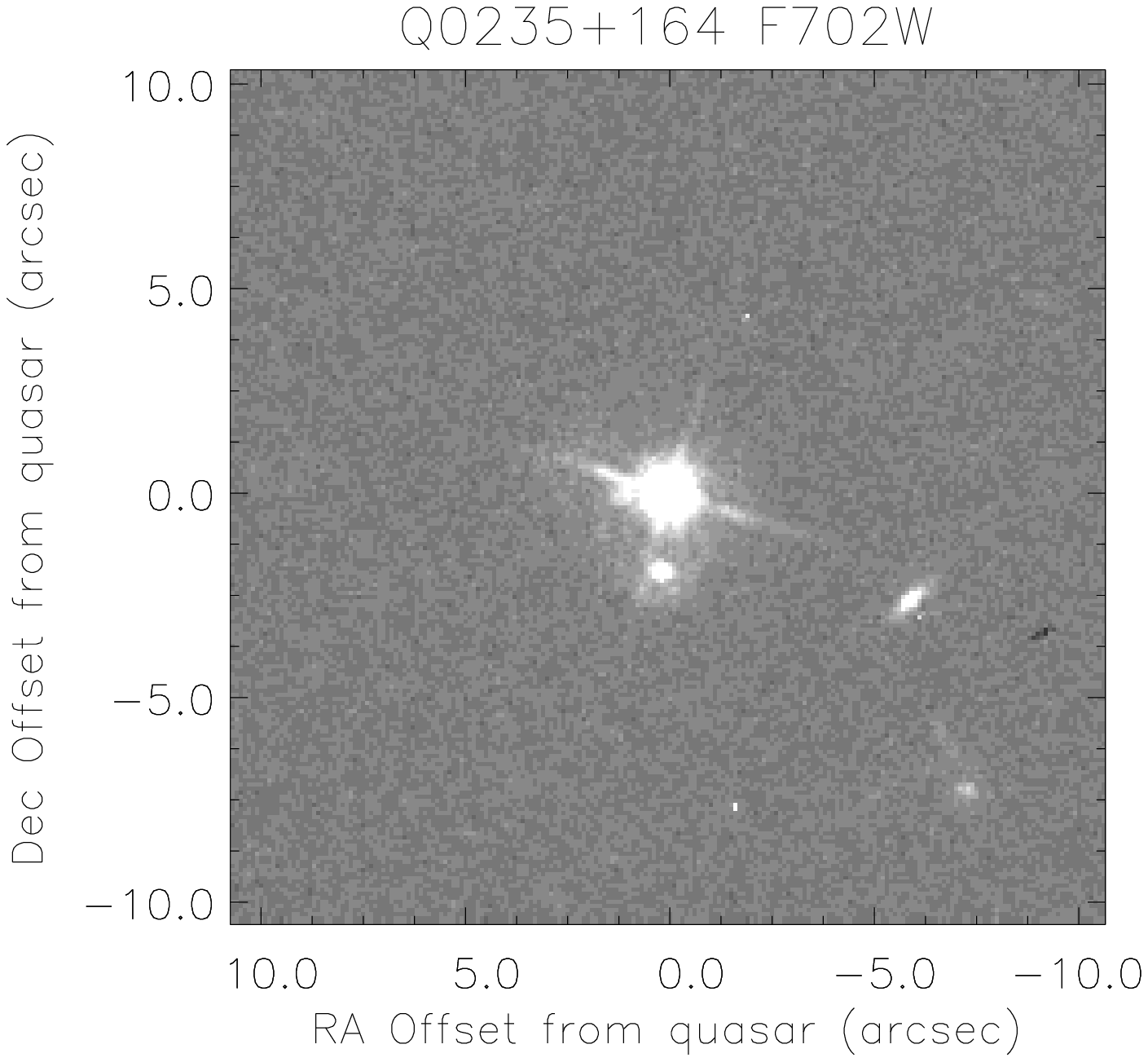}
\includegraphics[width=0.5\textwidth]{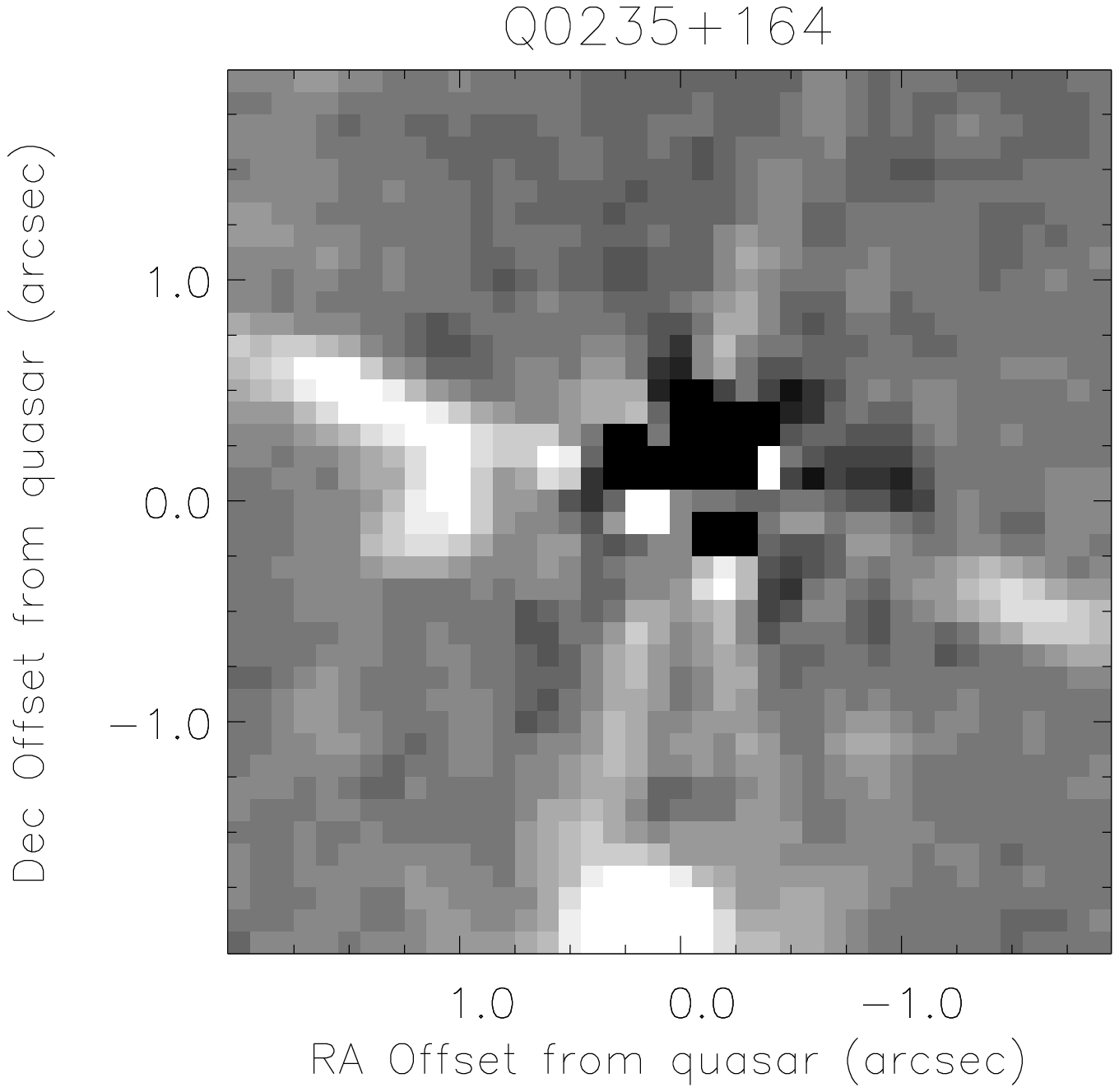}
\end{tabular}

\caption{ Upper left panel shows our Gemini/Hokupa$'$a H-band image smoothed
by a guassian of FWHM=0\farcs2.  The upper-right  panel
shows our Quasar-subtracted H-band image of the central $4 \arcsec$  smoothed
by a 3-pixel gaussian.  Contours overlaid on this image are 3$\sigma$ above the
sky in the unsmoothed image.  Lower panels shows the HST/WPFC2 F702W image 
for comparison.  The lower-left panel shows the HST image corresponding to our
full H-band image while the lower-right panel shows the azimuthal-average PSF 
subtracted HST image.  All figures are shown with a linear
intensity scale with North up and East to the left.  We identify object
+1.11-0.01 as the most likely candidate absorber.  This object
has been previously identified as a candidate absorber (e.g.  object A1 in
\cite[]{Yanny89}).  The object seen 2$\arcsec$~ south of the quasar is the BALQSO
object (Object ``A'' identified by \cite[]{Burbidge96}).  There is a small
extension SW of the center of the field
but it is too close to the center of the field to be distinguished  from
residuals of the quasar subtraction.\label{fig:0235+164}}

\label{fig2}
\end{figure}

Figures~\ref{fig:0054+145}-\ref{fig:1329+412} show the reduced image and the 
QSO subtracted image for each field.   For fields  where HST archival
imaging data are available, we show the HST images for  comparison.  These
archival data are from the HST programs of Bahcall (proposal ID: 5343, Q0054+144),
Burbidge (5096, Q0235+164) , Lanzetta (5949, Q0850+440), 
Bechtold (9173, Q1127-145), and Steidel (5984, Q1329+412).

\section{ANALYSIS}
\subsection{OBJECT DETECTION}

We  used the  automated software SExtractor~\cite[]{Bertin96} to detect sources
in our images. We ran SExtractor on the regions with the highest
signal-to-noise ratios of the azimuthally-averaged QSO-subtracted images. These
regions spanned approximately $15\arcsec$ on each side.  In addition, we ran
SExtractor on the regions centered at the position of the quasar in  the KL
quasar-subtracted images. These smaller regions measured $4\arcsec \times
4\arcsec$.  To look for extremely faint objects, we ran SExtractor on the
smoothed  images.  In this latter case, we used the azimuthally-averaged quasar
subtracted images and convolved the images with a gaussian with a FWHM of
20~pixels (0\farcs4). 

SExtractor requires a number of input parameters to work properly. Naturally,
the driving parameters are the  detection threshold and the number of connected
pixels above that threshold. After many test runs,  we chose a detection
threshold per pixel of $1\sigma$ of the sky level and a minimum number
of contiguous pixels of 75. Connected pixels are defined  as
pixels touching  any of their sides or corners  as implemented in SExtractor. 
Thus for a detection to be triggered, there had to be 75 connected pixels each
with an intensity at least $1\sigma$ above the sky.   Our $1\sigma$ per  pixel
threshold  corresponds to a surface brightness of  $\mu_{1\sigma,lim}\sim
20.0$~mag/arcsec$^2$ in H and $\mu_{1\sigma,lim}\sim 19$~mag/arcsec$^2$ in K$'$
(Table~\ref{tab:obs}). In comparison, the sky brightness in $H$ at Mauna Kea is
$\mu_H\sim13.4$ mag/arcsec$^2$ \cite[]{Tokunaga99} and $\mu_{K'}\sim 12.8$
mag/arcsec$^2$ (Gemini Hokupa$'$a instrument web page) thus our limiting
surface brightness is $\sim 300-400$ times fainter than the sky.  

At these flux limits, we expect to be able to measure image structure for the
brighter disk galaxies and to detect dwarf galaxies.  Disk galaxies in the Coma
cluster have typically $\mu_e$\footnote{$\mu_e$ is the surface brightness at
the  half-light radius}$=17.2$  mag/arcsec$^2$ \cite{Gavazzi00} so that at the
target absorbers with $z\le0.2$, we expect disk galaxies to have $\mu_e$ in the
range 20.2 - 21.7 mag/arcsec$^2$.  These are just at the surface-brightness
limit of our data.  Our object detection threshold is  much fainter.  For
example, Virgo cluster dwarf galaxies have a median total H $\sim 12$
mag~\cite[]{Gavazzi01} and would measure $H\sim$20 mag at the redshifts of our
sample, assuming  a flat spectrum.  The limiting magnitude of our observations,
defined as the magnitude corresponding to a $3\sigma$ flux within an aperture
of 75 pixels, is approximately $H_{3\sigma,lim}\sim23.3$ and
$K'_{3\sigma,lim}\sim21$ (see Table~\ref{tab:obs}).   At our flux limits we
easily detect typical dwarf galaxies at the redshifts of these absorbers. 

The detected sources are presented in Table~\ref{tab:obj}.  We detect a total
of 31 sources around 6 quasars.  The naming convention for objects corresponds
to the offset east and north in arcseconds from the quasar centroid. Each of
the six quasar fields contains at least one object and five of the six fields
show objects previously undetected.

\subsection{Photometry} 

Absolute photometry is based on observations of  standard stars taken during
the same night before and after the observations of the  quasar field when
observing conditions were photometric. For data obtained under  non-photometric
conditions, we used the quasar and its 2MASS catalog magnitude to  boot-strap
the photometry of the objects in the fields.   This is reasonable as long as
the quasar is not highly variable in the near-infrared.  There are two cases
when the night was not photometric and the quasar is known  to be highly
variable.  For Q0235+164 we report relative photometry only.  In the case
of Q1127-145, we used the photometry of object G1 from ~\cite{Chen01} kindly
provided to us by Hsiao-Wen Chen to boot-strap the photometry for the rest
of the objects in the field. 

All photometry is reported here as AB magnitudes.   The following relationship
between the Vega and AB systems \cite[]{OkeGunn83} was used:
\begin{equation}
m_{AB} = m_{Vega} - 2.5~log (f_{\nu,0}) + 8.90 \label{AB}
\end{equation}
where the zero-point flux is taken to be $f_{\nu,0} = 1080$ Jy in H and 
$f_{\nu,0} = 670$ Jy in K$'$ \cite[]{Tokunaga99}.  
In Table~\ref{tab:obj}, we list for  each detected object the ID (right
ascension and declination offsets relative to the quasar); the angular distance to the QSO, $\Delta
\theta$ in arcsec, and the AB magnitude in the H or K$'$ band as determined
by SExtractor. These are the magnitudes determined using a metric 
radius defined as 2.5 times the first moment of the light profile
\cite[]{Kron80}: $r_{metric} = 2.5 \times {\int I(r) r dr}/ {\int I(r) dr}$
where $I(r)$ is the intensity profile as a function of radius $r$.

\subsection{Image Structure}

The image properties of the detected sources allow us to charaterize their
morphologies and to address the question of whether they are  bulge-dominated
or disk-dominated.  We determined the structural parameters for roughly half of
the detected sources by running GIM2D \cite[]{Simard02} on the
azimuthal-average subtracted images.  Out of the 31 sources detected, we were
able to  extract image structural parameters for nineteen sources.  Fourteen
sources had high total signal-to-noise ratio and are well separated from the
quasar. For five of the sources, the structural parameters are largely
uncertain (as seen by the errors in  Table~\ref{tab:morph}).  The remaining
objects were either too faint or too close to the quasar line of sight to
estimate  any image structural parameters.  In the end,  we are able to
constrain the morphologies of twelve out of the 31 sources. 

GIM2D fits two dimensional intensity profiles with a combination  of  the
S\'ersic law and  an exponential disk.  
%
%
Each source was fit with three types of models: (1) single
$r^{1/n}$-laws, (2) exponential disks plus $r^{1/4}$-laws, and (3) exponential
disks plus $r^{1/n}$ laws.   In all cases the PSF provided to GIM2D was the
azimuthally-averaged quasar image.

The output quantities of GIM2D include the total luminosity,  bulge-to-total
ratio ($B/T$), effective radius ($r_e$), ellipticity of the bulge, position
angle of the bulge, exponential disk scale length ($r_d$),  inclination angle
of the disk, position angle of the disk, x and y position offsets,  background
level, reduced-$\chi^{2}$ value, half-light radius, and S\'ersic exponent $n$. 
These values are provided with their 99\% confidence limit.  If the
distributions were normal, they would correspond to a $3\sigma$ error. We
assign a morphological class based on the values of $B/T$, $r_e$, $r_d$, and
$n$  and their 99\% confidence limits.   Table~\ref{tab:morph} presents the
derived image structural parameters for each object.  We list the objects' ID,
the  bulge/disk ratio ($B/T$), the  scale lengths ($r_e$ and $r_d$) in
arcseconds, and the exponent $n$ of the generalized $r^{1/n}$-law. For the
bulge-to-total ratio, scale lengths, and exponent, we also present the 99\%
confidence limits obtained with GIM2D.

An important motivation for the study was to determine whether bulges or disks
dominated galaxies account for the low-redshift absorbers.   As such, in
assigning  a morphological type,  we identified the objects as point sources,
disk-dominated, bulge-dominated, or a combination of disk and bulge.   Of the
sources we could measure the structural parameters, 3 of them have FWHM that
are identical to the PSF.  For these sources, GIM2D returned zero scale lengths
and we interpret these as point sources.  Single $r^{1/n}$ fits were used to
determine if any of the sources were pure disks.  A pure disk would have a
value of $n=1$ in the S\'ersic-law only fits.   We found that only two sources
are consistent with being pure disks. We found $n=0.96^{+0.75}_{-0.05}$ for
Q0235+164~-5.85-2.69 and  $n=1.43^{+0.12}_{-0.20}$ for Q1127-145~+14.5-6.76.
These values are also consistent with the fits done with two components (see
Table~\ref{tab:morph}) where they have a small bulge ($B/T<0.27$) and large
disk scale length ($r_d\sim 0\farcs2$).  

The outputs of the $r^{1/4}$ and the generalized  $r^{1/n}$ fits are consistent
with each other within the confidence limits.  By and large we found that the
$r^{1/n}$+disk fits showed a smaller residual signal and we therefore provide
the results of these decomposition instead of that of the $r^{1/4}$+disk. 
Even  though we allowed n to vary, 10 of the sources returned $n\sim4$ as the
best fit.  We used the reduced-$\chi^2$ values that are output by GIM2D to discard any
bad fits.  There are only two cases in which the reduced-$\chi^2$ values are large. From
the two component fits, we find that there are 6 disk-dominated galaxies, 3
disk+bulge galaxies,  2 bulge-dominated galaxies, 3 point sources, and 5
unconstrained objects.

\section{RESULTS}

Below we present  the astrometric, photometric, and morphological results
derived for the features in each field.  Table 5 summarizes the morphologies,
impact parameters, luminosities, and  scale lengths for the candidate objects,
assuming that they are at the redshifts of the absorbers.  The morphology is
not listed for the objects that are  too faint or too close to another object,
since the profile fits for these objects  are not robust. The scale lengths of
the profiles are not listed for the point sources and for the sources where the
profiles could not be fit. For candidate absorbers, we converted  fluxes to
luminosities using $L = 4\pi d_L^2 F$ where $d_L$ is the luminosity distance of 
the DLA candidate.  These luminosities are express units of $L^{*}$ where we
have adopted 
$L_{H}^{*} = 1.33 \times 10^{43}$ erg $s^{-1}$
(\cite{Kulkarni00}) and  $L_{K}^{*} = 3.62 \times 10^{42}$ erg $s^{-1}$
(\cite{Bell2003}).

\subsection{Q0054+144}

Q0054+144 is a radio-quiet X-ray-bright QSO at a redshift of $z_{em} = 0.171$. 
This object was imaged with HST WFPC2 (Bahcall et al. 1996; McLure et al.
1999).  These HST data indicate that the host galaxy is well described by an 
early-type galaxy. A DLA candidate absorber at $z = 0.103$ with a neutral
hydrogen column density of $\log N$(H~I) = 20.1 was suggested by  Lanzetta et
al. (1995) using IUE data. However, higher resolution HST  GHRS spectra showed
that no DLA absorption is present at $z = 0.103$ (Bechtold  et al. 2001) and no
X-ray absorption was detected in Chandra observations   of Bechtold et al.
(2001). A low-ionization metal absorption line system is  present at $z =
0.103$, but the Ly-$\alpha$ line is not damped, with   $\log N$(H~I)$ = 18.3$
(Turnshek \& Rao 2002).  Thus this is a Lyman-limit system.  

Figure \ref{fig:0054+145} shows our H-band image of the field, the
quasar-subtracted image of the central portion of the field,  and the HST/WFPC2
F606W image. Our H-band image is $20 \arcsec \times 20 \arcsec$  which
corresponds to  $\approx 40 \times 40$ kpc$^{2}$ at $z = 0.103$  .  We detect
an object approximately 0\farcs8 SW of the quasar only after the KL QSO
subtraction.  It is too close and too faint to measure its magnitude or
structural parameters.  It is in the same direction and approximately the same
location as the object identified in \cite[]{McLure99} (See their figure A11)
but is considerably less extended.  This smaller extent may be due to the 
smaller region over which the quasar subtraction was applied with the KL 
technique.  In addition, a number of faint objects are
seen around this elliptical galaxy in both our image and the HST image.  They
are all small objects and could be companions to the host galaxy.  An
additional object lies approximately $12 \arcsec$ south of the quasar just at
the edge of our field of view and appears to correspond to an object in the HST
image of McLure et al. (1999). This object was excluded from our analysis due
to its close proximity to the edge of our field.

\subsection{Q0235+164}

Q0235+164 (AO 0235+164) is a radio-loud, optically violently variable, X-ray and 
$\gamma$-ray emitting blazar. Roberts et al. (1976) measured a complex 21-cm 
absorption profile in the radio spectrum of AO centered at z = 0.524.  Based on
a UV spectrum of the QSO obtained with HST/STIS,  Cohen et al. (1999) confirmed
that the absorber is a DLA  with $N_{\rm H I} \approx 5 \times 10^{21}$
cm$^{-2}$. Junkkarinen et al. (2004)  have detected the 2175 {\AA} feature and
diffuse interstellar bands at the  redshift of this absorber. Two faint 
objects with [OII] 3727 emission, have been detected within 2\arcsec~ from AO, 
and have been suggested as possible sites for the z = 0.524 absorption  system
(Smith et al, 1977; Yanny et al, 1989). Burbidge et al. (1996)  using HST/WFPC2
and HST/FOS resolved these two objects more clearly.  The nearest one
(\cite{Yanny89}'s A1) might contribute to the complex H I 21-cm absorption, 
while the object 2\arcsec~ south of AO (\cite{Yanny89}'s A) is an AGN surrounded
by faint nebulosity which can be classified as a BALQSO.   In the optical and 
infrared imaging observations of the QSO, Chen \& Lanzetta (2003) concluded
that there is a group of galaxies  at the redshift of the known DLA several of
which likely contribute to  the DLA system. They find that the photometric
redshift for the object $2\arcsec$ south of the QSO is consistent with the
spectroscopic redshift of  the known DLA.   \cite{Yanny89} found [O~II]
emission from both the A1 and A objects.  

Figure \ref{fig:0235+164} shows our H-band image of the field before and  after
subtraction of the quasar image as well as the archival HST/WFPC2 F702W image.
The full images are $20 \arcsec \times 20 \arcsec$  corresponding to $\approx
120 \times 120$ kpc$^{2}$ at $z = 0.524$.   We identify six objects.  The
angular distances of these  objects to the QSO range from $\Delta \theta = 0.6
\arcsec$ to $10.2 \arcsec$  corresponding to $b_{abs} = 3$  to $60$ kpc.  
Objects other than object -4.93-7.47 have already been reported in the
literature.  It is unlikely to be the absorber since at the redshift of the
absorber it would have an impact  parameter of more than 50 kpc and its
profile scale length is close to the FWHM of the quasar.  Object -7.15-7.21
appears to be  a point-source in the HST images but shows a linear extension in
both our H-band image and the HST image.  Object +0.15-1.91 is the BALQSO
object found by Burbidge et al. (1996).   Its morphology in the NIR is extended
and disk dominated.    An object is detected by SExtractor in the PSF
subtracted image at a separation of 0.5\arcsec~ from the quasar centroid
(-0.3-0.4).  We  have disregarded this object since it falls within the region
where PSF residuals are seen in PSF subtractions of stellar images.

We regard +1.11-0.01 (object A1 in Burbidge et al. (1996)) as the likely 
absorber.  Its photometric redshift is consistent with it being at the absorber
redshift \cite[]{Chen03} and its profile is consistent with a combination
exponential disk and $r^{1/4}$.  Absolute photometry was not possible from our 
data for this field  because of the observations were made under
nonphotometric  conditions.  SExtractor identifies another object 2\farcs5 NE
of the quasar.  This  object was not reported in previous studies.  While
SExtractor identifies this and object +1.11-0.01 as separate objects, a 1-D cut
across these two objects is well fit by a bulge+disk profile. If these are the
same object, then it could be a spiral galaxy slightly  inclined to our line of
sight.  At the redshift of the absorber, it would have an impact parameter of
about 6-7 kpc.  We note that if the object +2.40+0.93 is an extension of object
+1.11-0.01, then the extension is in the NE direction. This is perpendicular to
the orientation suggested by \cite{Burbidge96}.

\subsection{Q0738+313}

Q0738+313 (OI 363) is a core-dominated slightly variable quasar at  $z_{em} =
0.635$. Rao et al. (1998) reported the discovery of two DLA systems  toward the
quasar at $z = 0.0912$ and $z = 0.2212$ with $N(H~I) =  (1.5 \pm 0.2) \times
10^{21}$   and $(7.9 \pm 1.4)\times 10^{20}$ cm$^{-2}$, respectively.  They
concluded that a  galaxy at $5.7 \arcsec$   from the QSO line of sight is the
only reasonable candidate at either  absorption redshift. \cite{Cohen01}
reported galaxies at $z = 0.221$ and  $z = 0.106$, $6 \arcsec$ and $28\arcsec$
away from the quasar line of sight, respectively.  She  suggested that the $z =
0.106$ galaxy may be a member of a cluster that causes the absorption at $z =
0.0912$. The morphology of this galaxy was classified as early-type by Rao et
al. (1998). Optical and infrared imaging observations  of the QSO made by
Turnshek et al. (2001) indicate that the DLA galaxy at  $z = 0.2212$ is a
``faint neutral colored galaxy with dwarf galaxy-like K and B-band
luminosities.'' Its spectrum is that of an early-type galaxy.  Turnshek et al.
(2001) also suggested that the putative $z = 0.0912$ DLA galaxy  is likely to
be all or part of the resolved light surrounding the quasar with armlike and 
jetlike-features. They suggested that the DLA is a low  surface brightness
dwarf galaxy, possibly an irregular or interacting  system.

Six objects are detected in our image.  Five have previously been
identified.  Object +1.90-5.38, the dwarf galaxy at $z = 0.221$ and
designated ``G1'' by others \cite[]{Cohen01,Turnshek01} is a disk-dominated
galaxy with a bulge-to-total ratio of 0.34.  This is consistent with an 
E/S0 galaxy suggested by \cite{Turnshek01}.  The object +2.02+1.54, 
designated ``S1'' by \cite{Turnshek01}, is still unresolved in our images, 
and GIM2D identifies the object as a point-source.

\begin{figure}
\begin{tabular}{cc}
\includegraphics[width=0.5\textwidth]{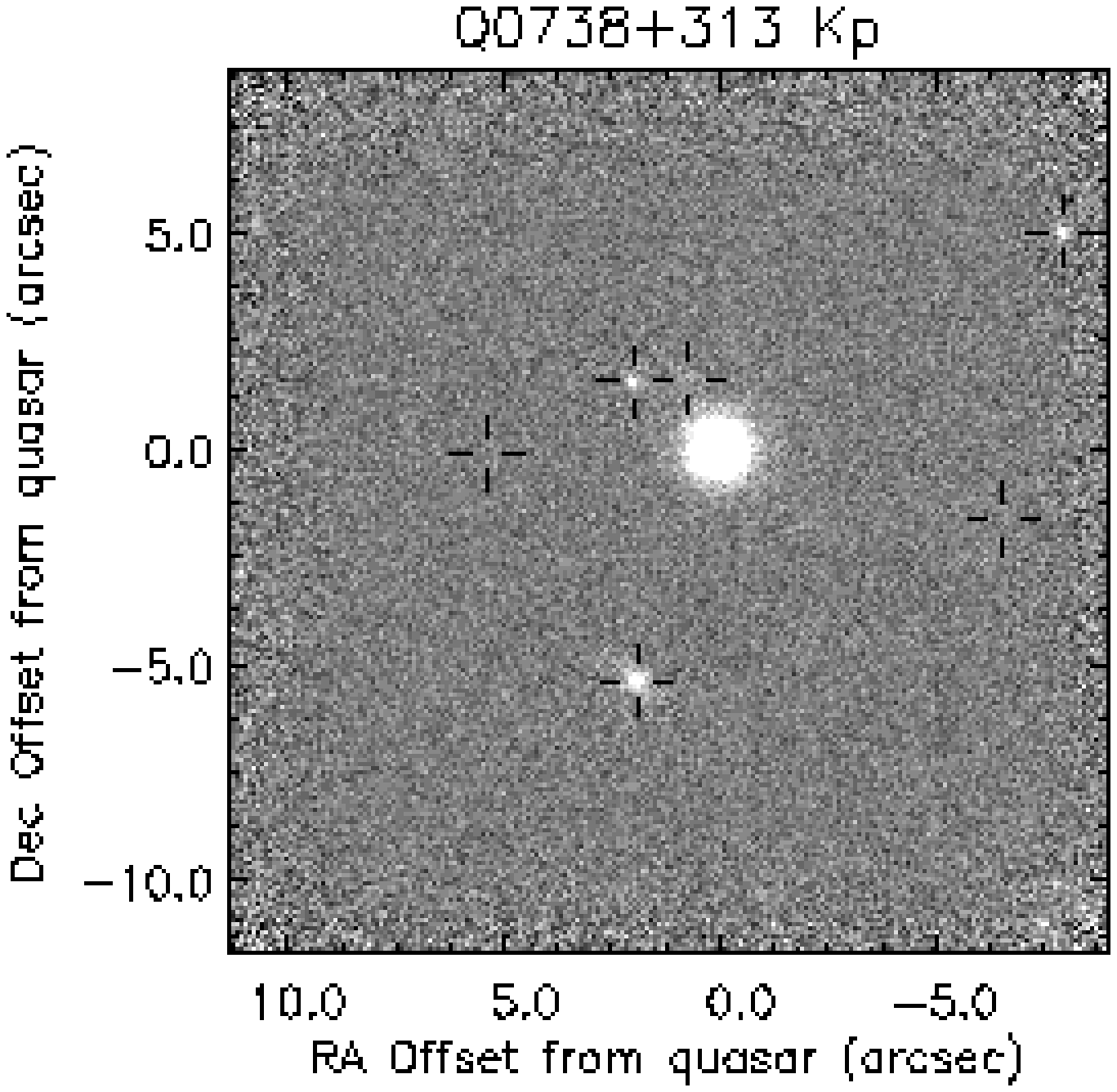}
\includegraphics[width=0.5\textwidth]{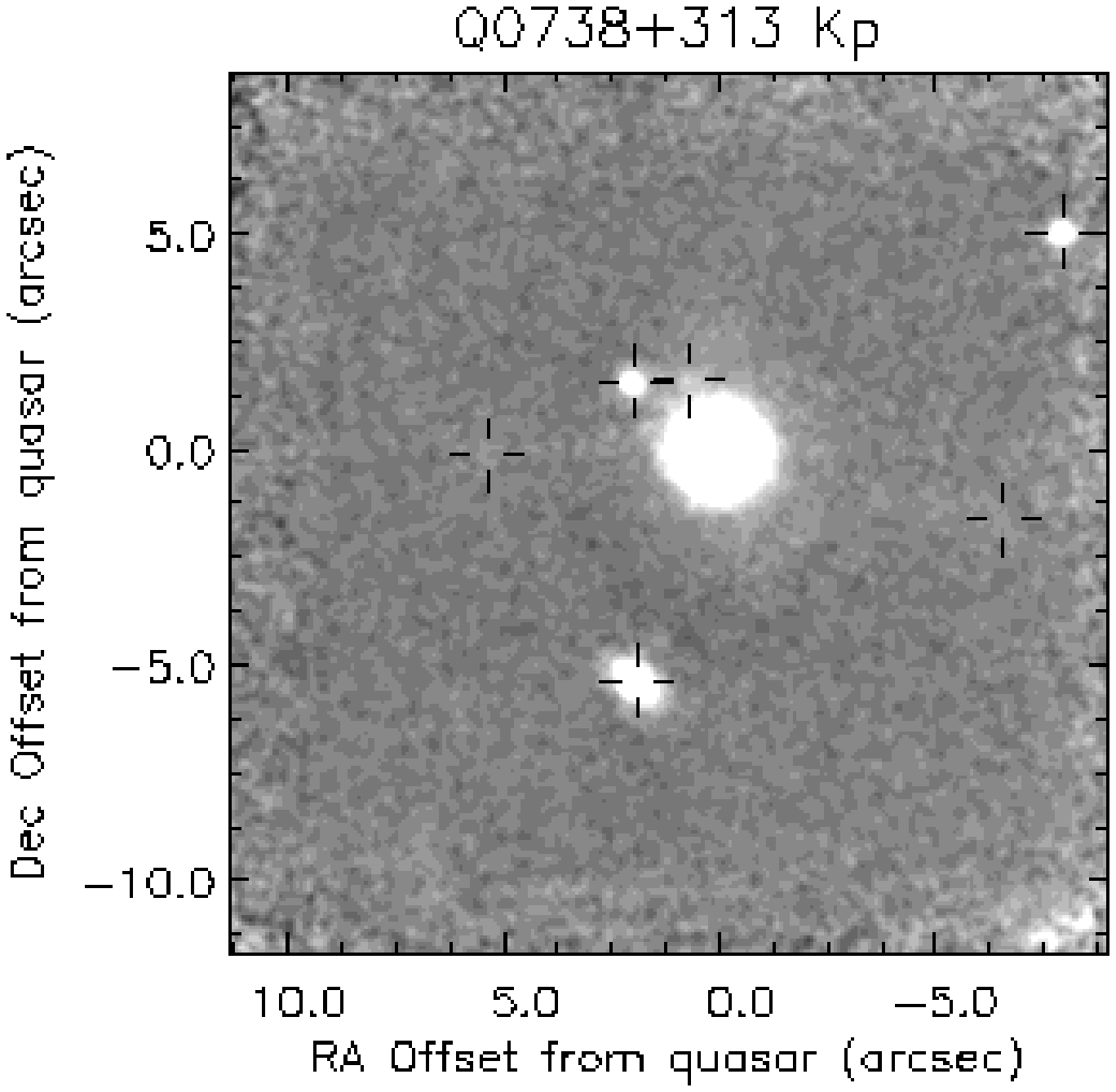}\\
\includegraphics[width=0.5\textwidth]{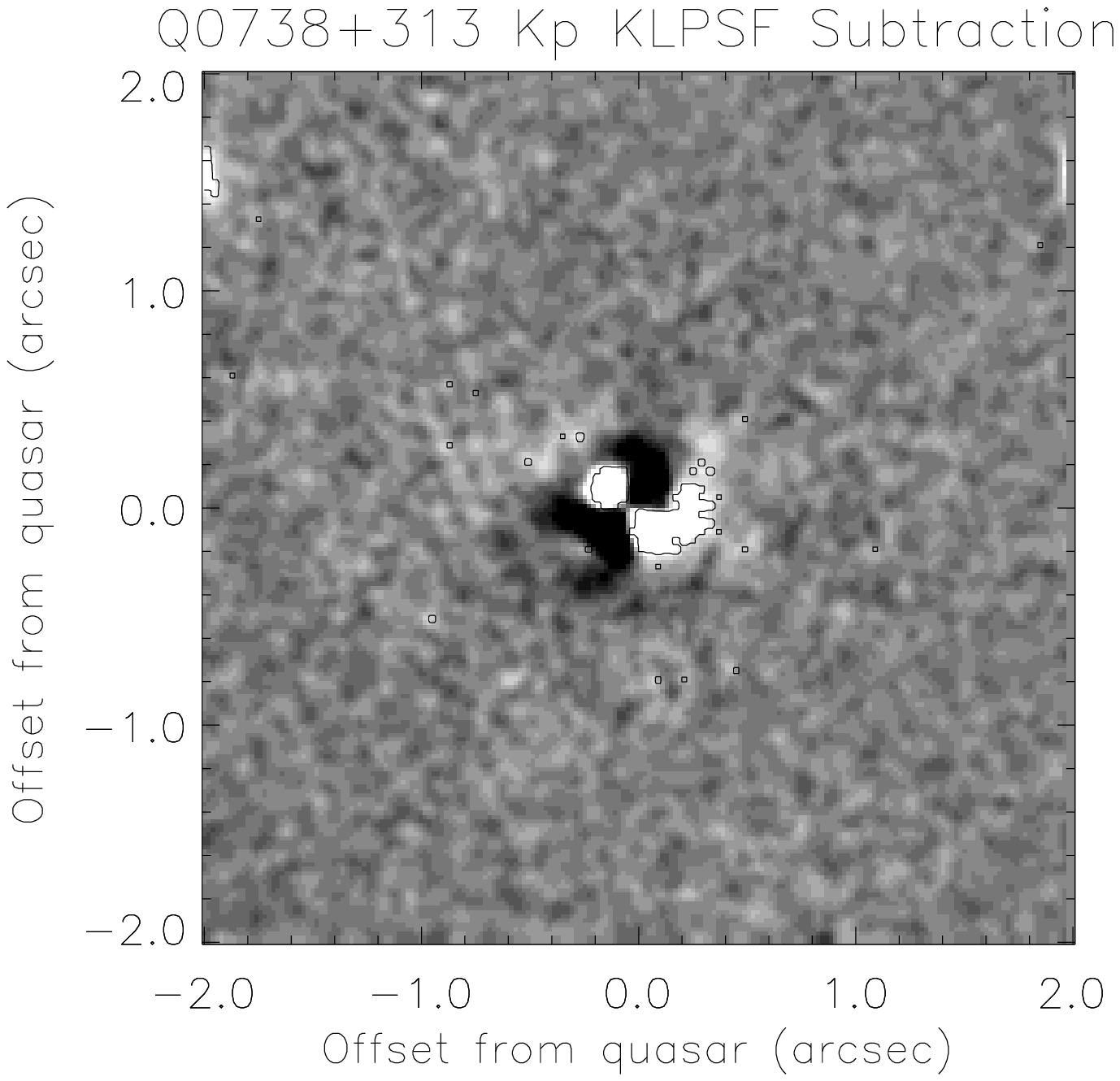}
\includegraphics[width=0.5\textwidth]{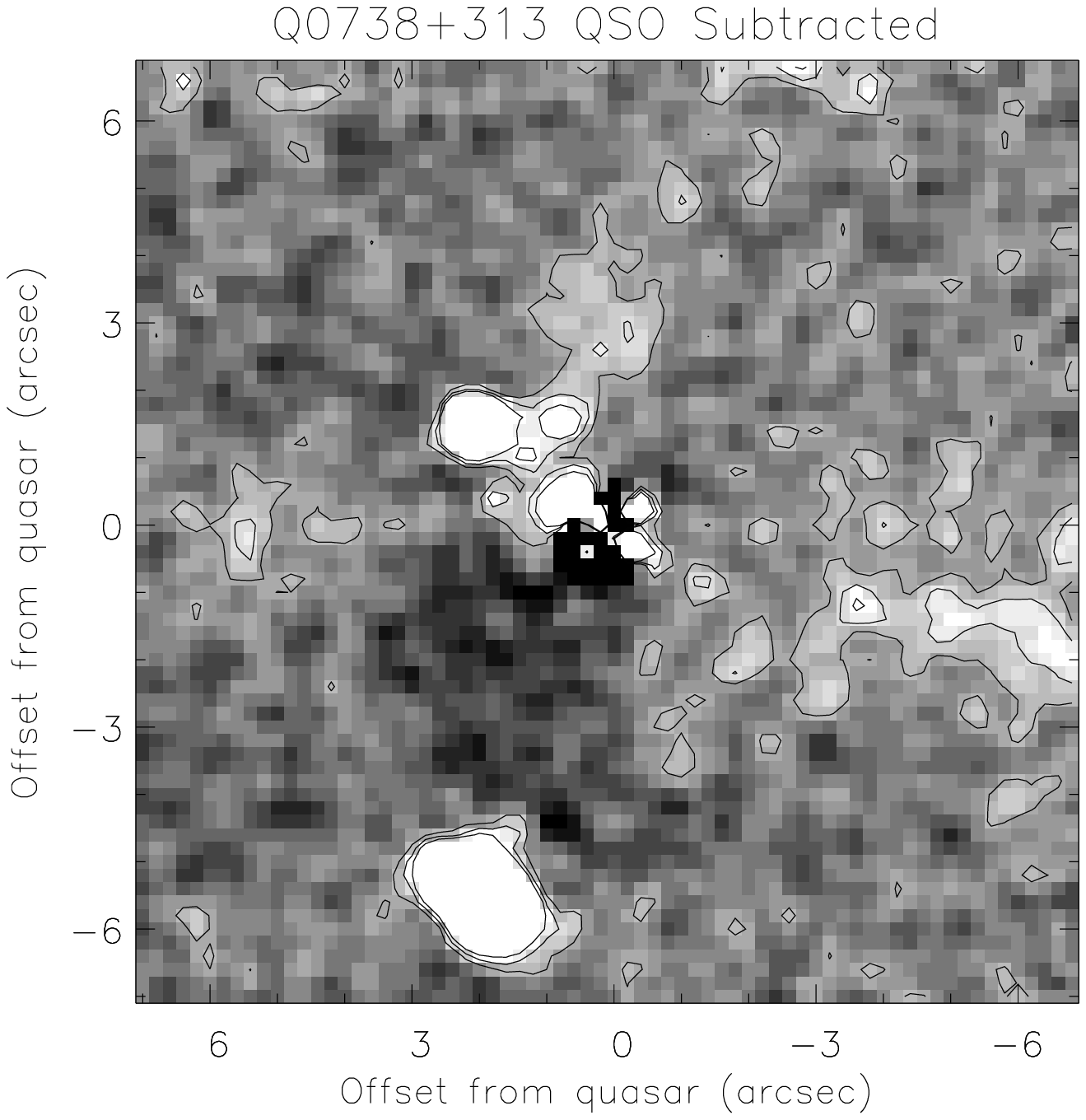}
\end{tabular}
\caption{
Upper panels show our Gemini/Hokupa$'$a K$'$ image of Q0738+313 unsmoothed (left) and 
smoothed (right) by a gaussian with FWHM=0\farcs2.   The full image in the 
top two panels are $20\arcsec$ on a side corresponding to 30 kpc at $z = 0.0912$ 
and 70 kpc at
$z = 0.2212$.  The lower-left panel  shows the KL QSO-subtracted image.  The 
lower-right figure shows the K$'$ image 
after smoothing by  a gaussian with FWHM equal to twice the FWHM of
the quasar image, subtracting an azimuthal averaged PSF, and then binning the
image to $0.2\arcsec$ pixels.    The image  shows the $14\arcsec \times
14\arcsec$ region centered on the location of the quasar.  The contours are 1-, 2-
and 3-sigma above the sky in the smoothed-rebinned image.  The one-sigma per
pixel level corresponds to 22.3 mag~arcsec$^{-2}$.  The 'jet' and 'arm' 
reported by \cite{Turnshek01} are apparent as are new features NNE of the 
subtracted quasar. \label{fig:0738+313}}
\label{fig3}
\end{figure}

\begin{figure}
\begin{tabular}{cc}
\includegraphics[width=0.5\textwidth]{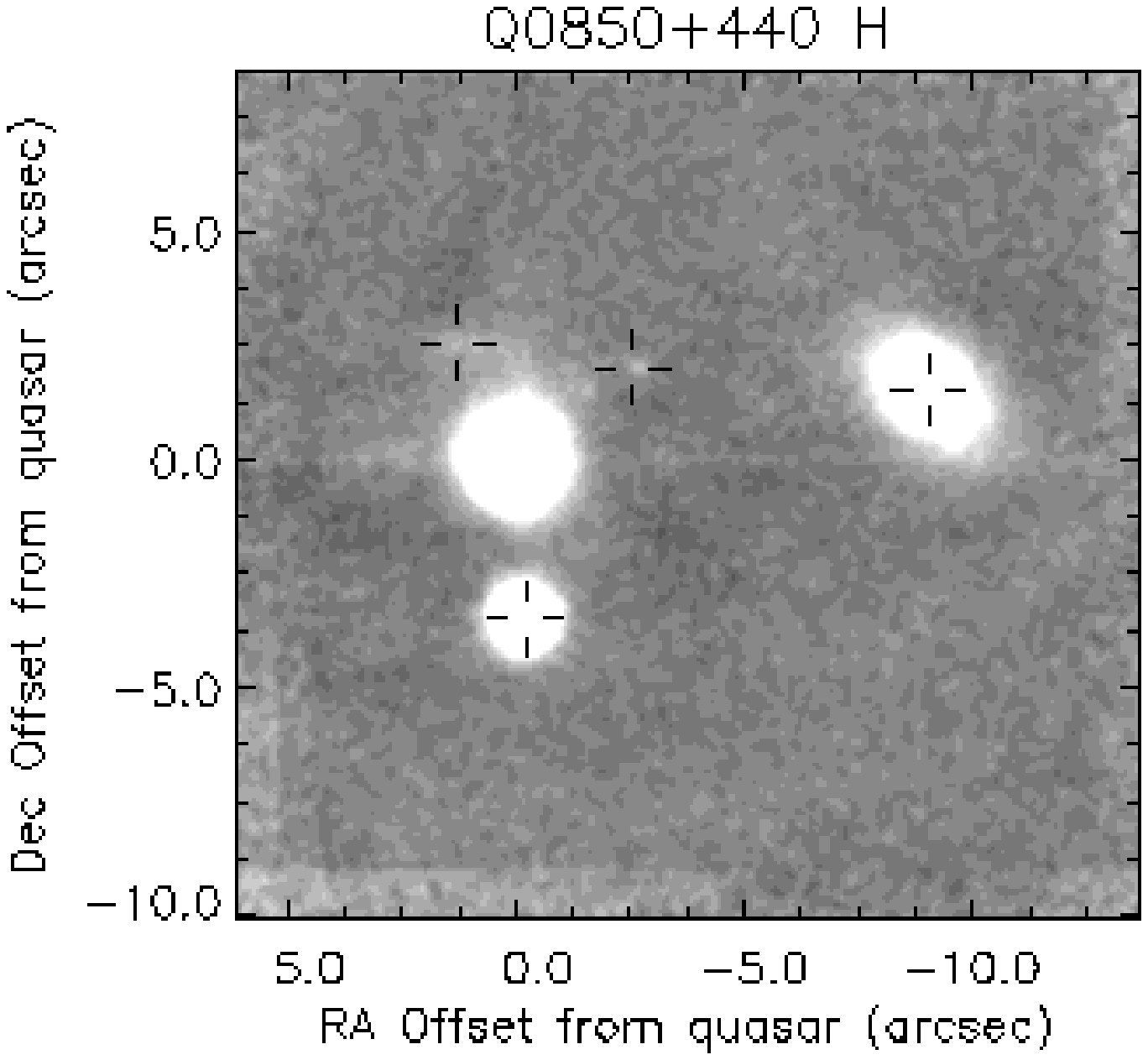}
\includegraphics[width=0.5\textwidth]{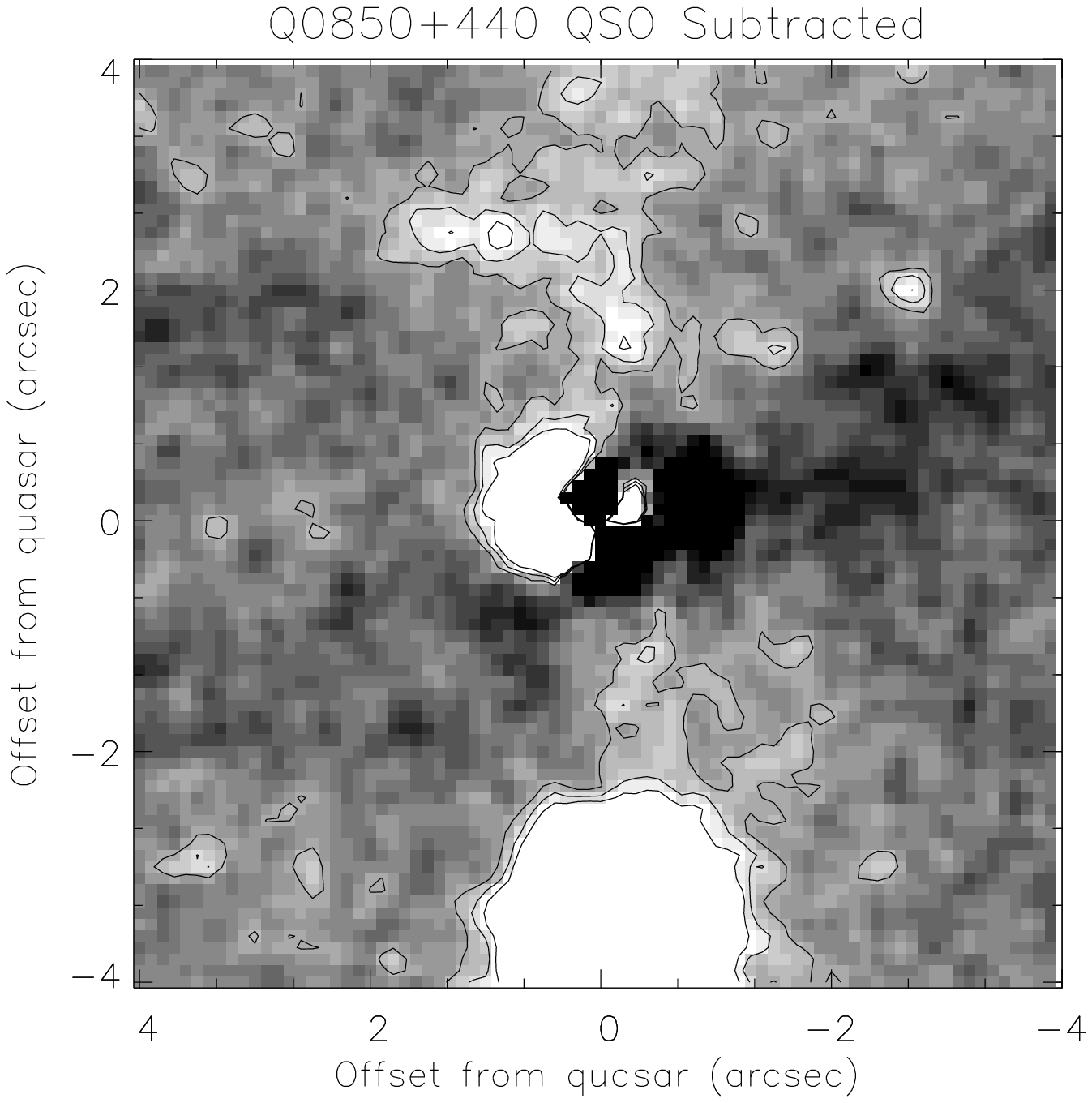}\\
\includegraphics[width=0.5\textwidth]{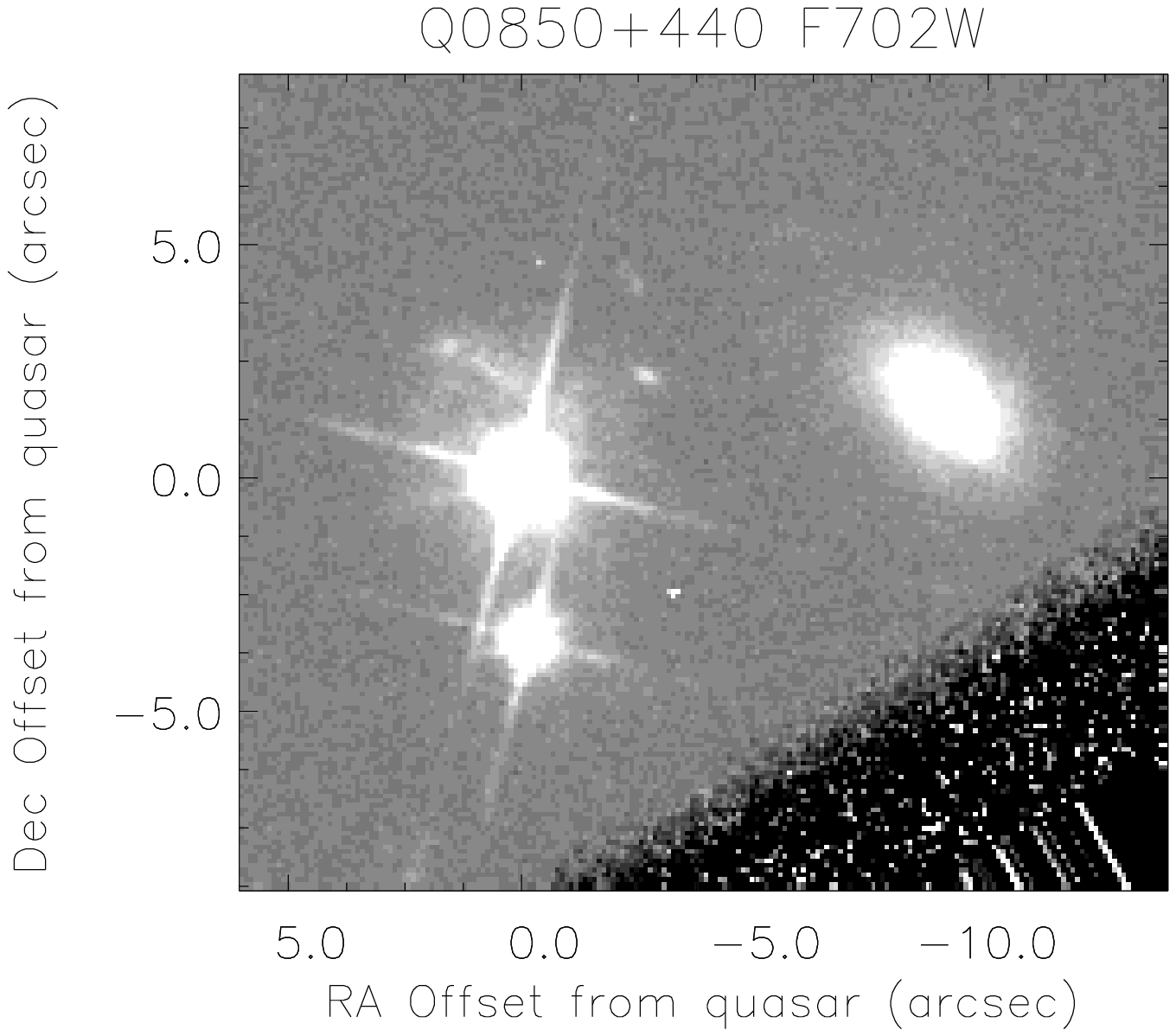}
\includegraphics[width=0.5\textwidth]{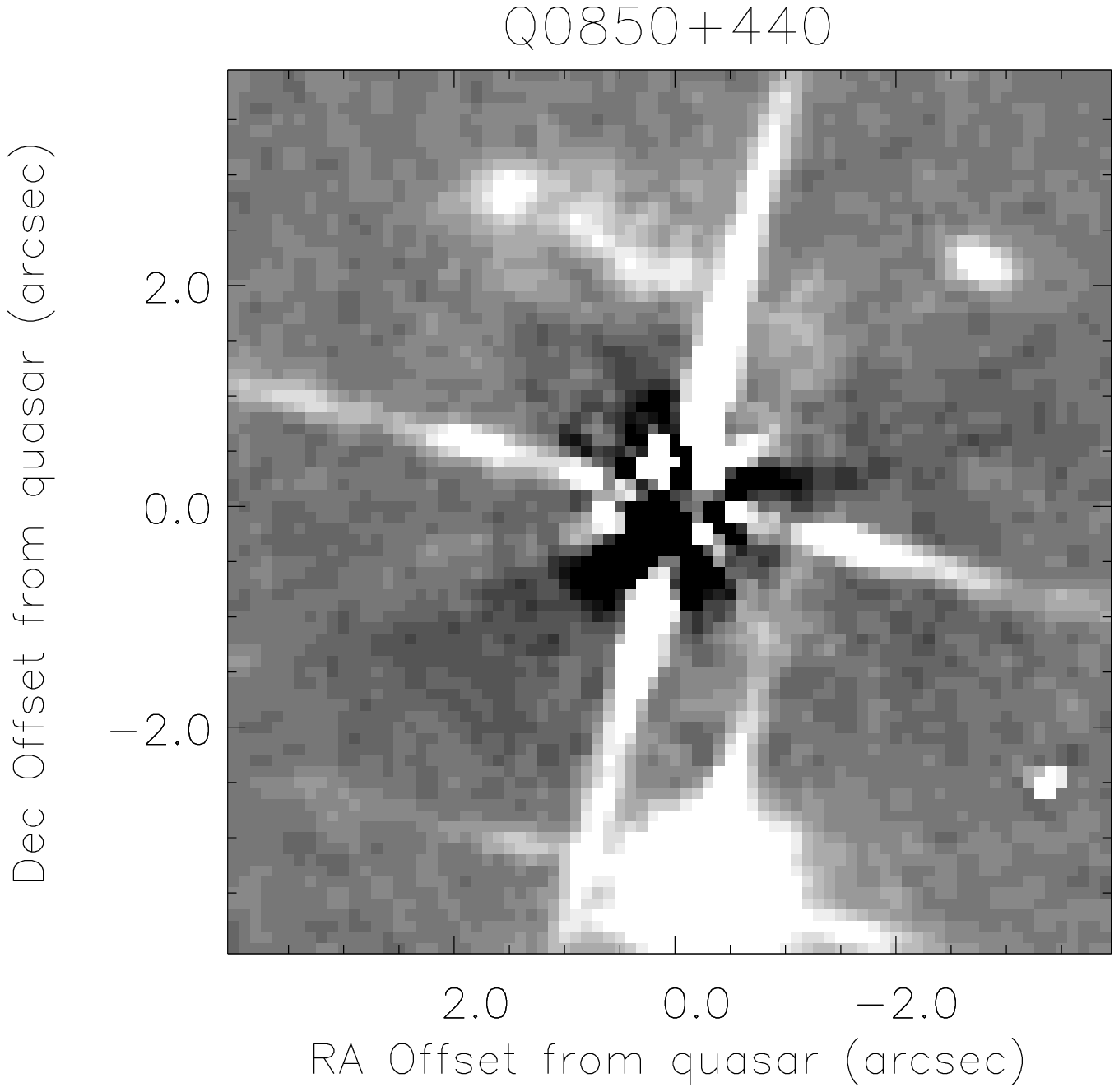}
\end{tabular}
\caption{
The upper panels show our H-band images of the field around the quasar before 
and after subtraction of  the PSF image. The full image (upper-left) 
corresponds to $\approx 54$ kpc at $z = 0.1638$.  We found objects close to 
the line of sight to the quasar with several arcsecond extensions.  The
extent of the emission is shown in the $8\arcsec \times 8\arcsec$ region 
centered on the azimuthal-average  QSO-subtracted image (upper-right).  
This image was smoothed by  a gaussian with  FWHM=0\farcs2.   
The contours are 1-, 2- and 3-sigma  above the sky in the 
smoothed-rebinned image.  The lower panels show the HST F702W image (left)  and
its azimuthal-averaged subtracted image (right).
}
\label{fig5}
\end{figure}

\begin{figure}
\begin{tabular}{cc}
\includegraphics[width=0.5\textwidth]{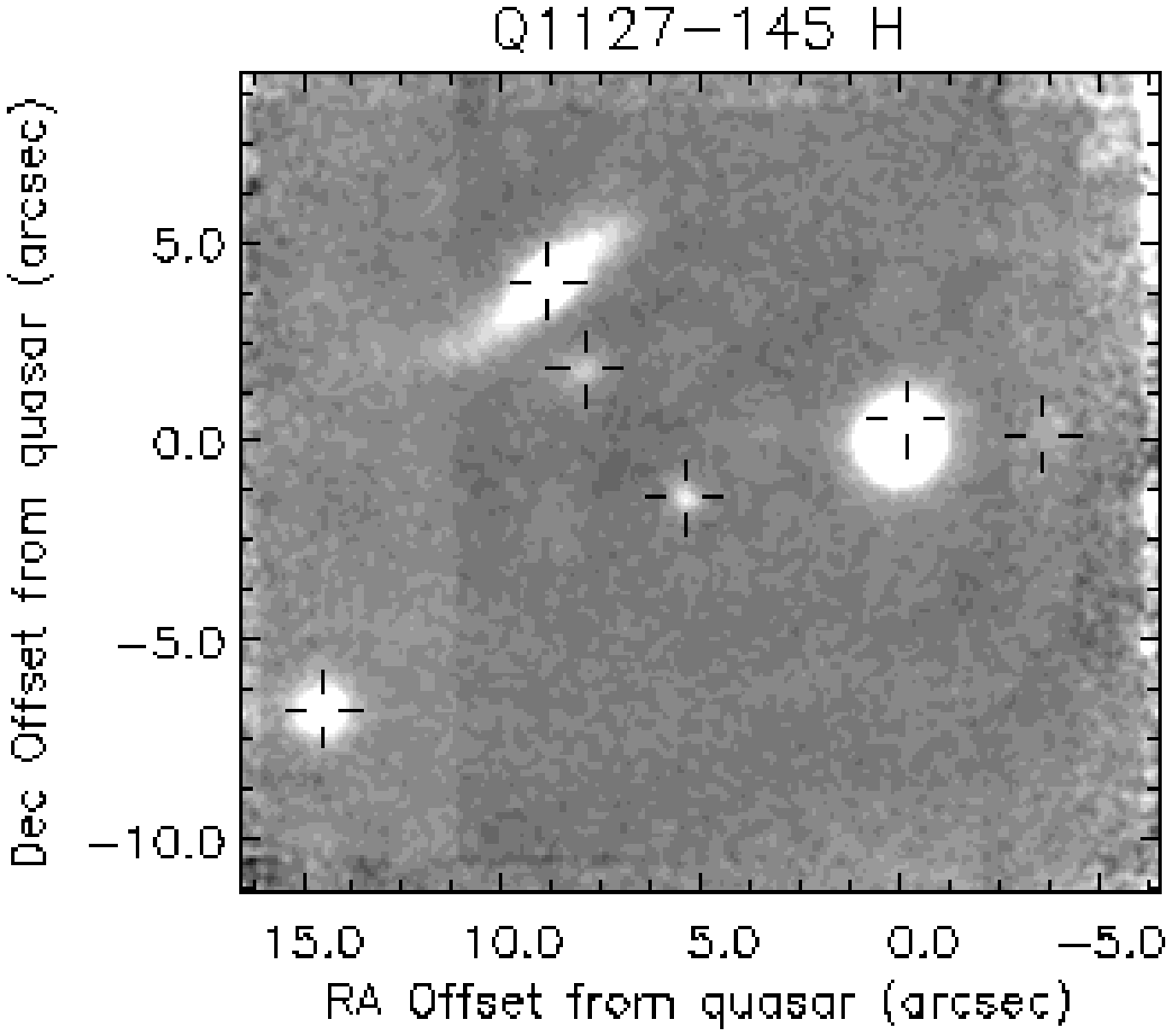}
\includegraphics[width=0.5\textwidth]{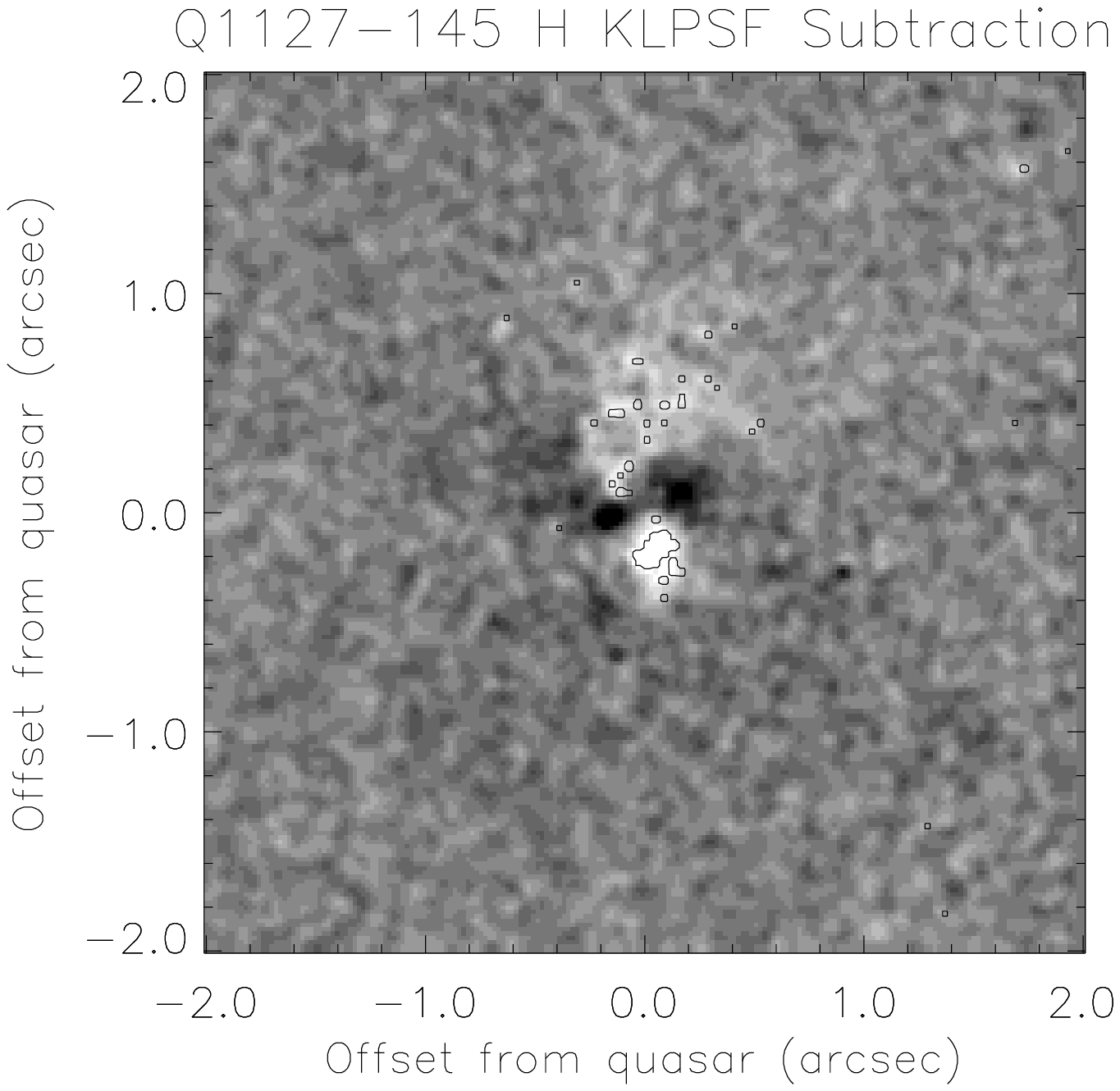}\\
\includegraphics[width=0.5\textwidth]{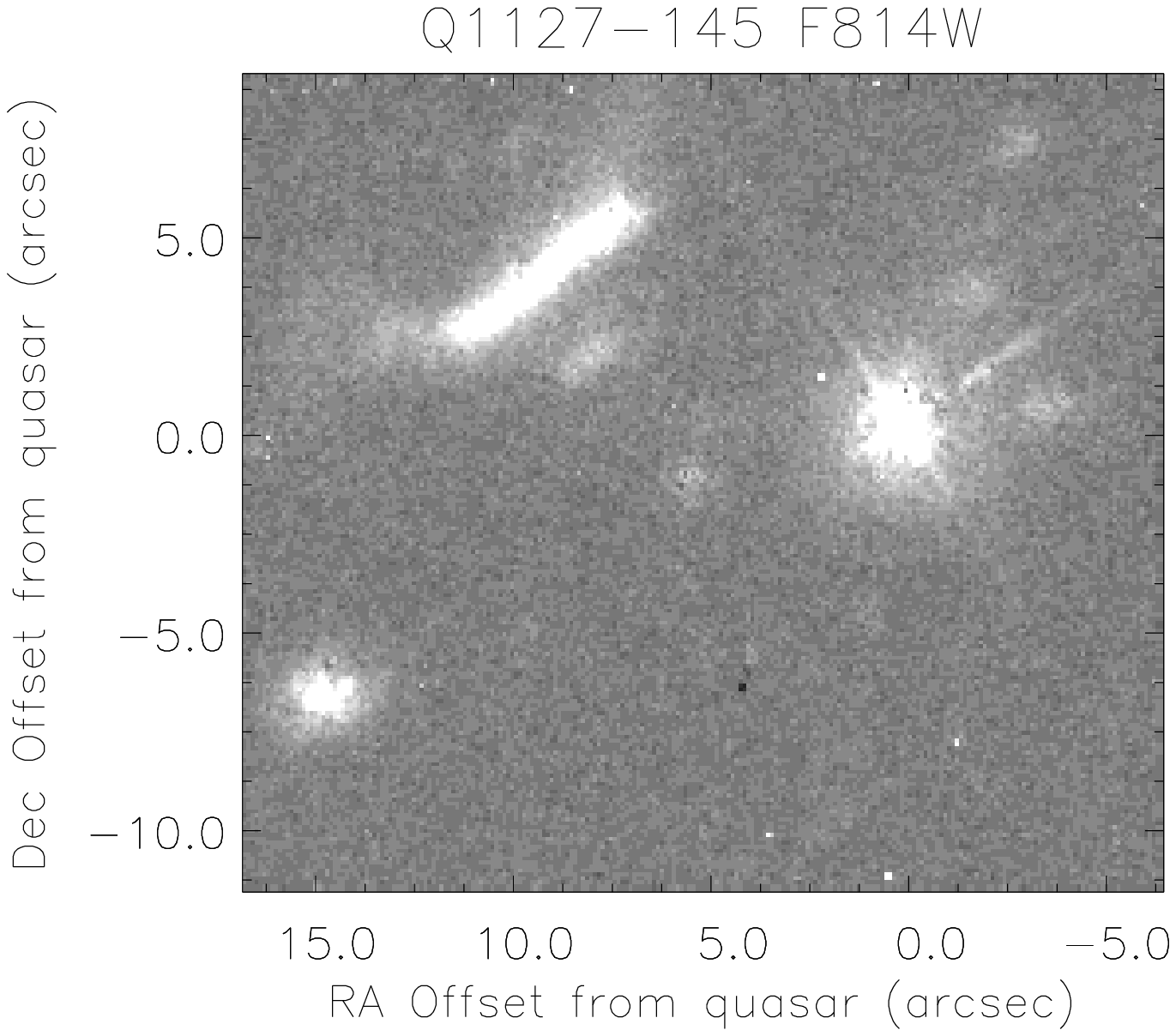}
\includegraphics[width=0.5\textwidth]{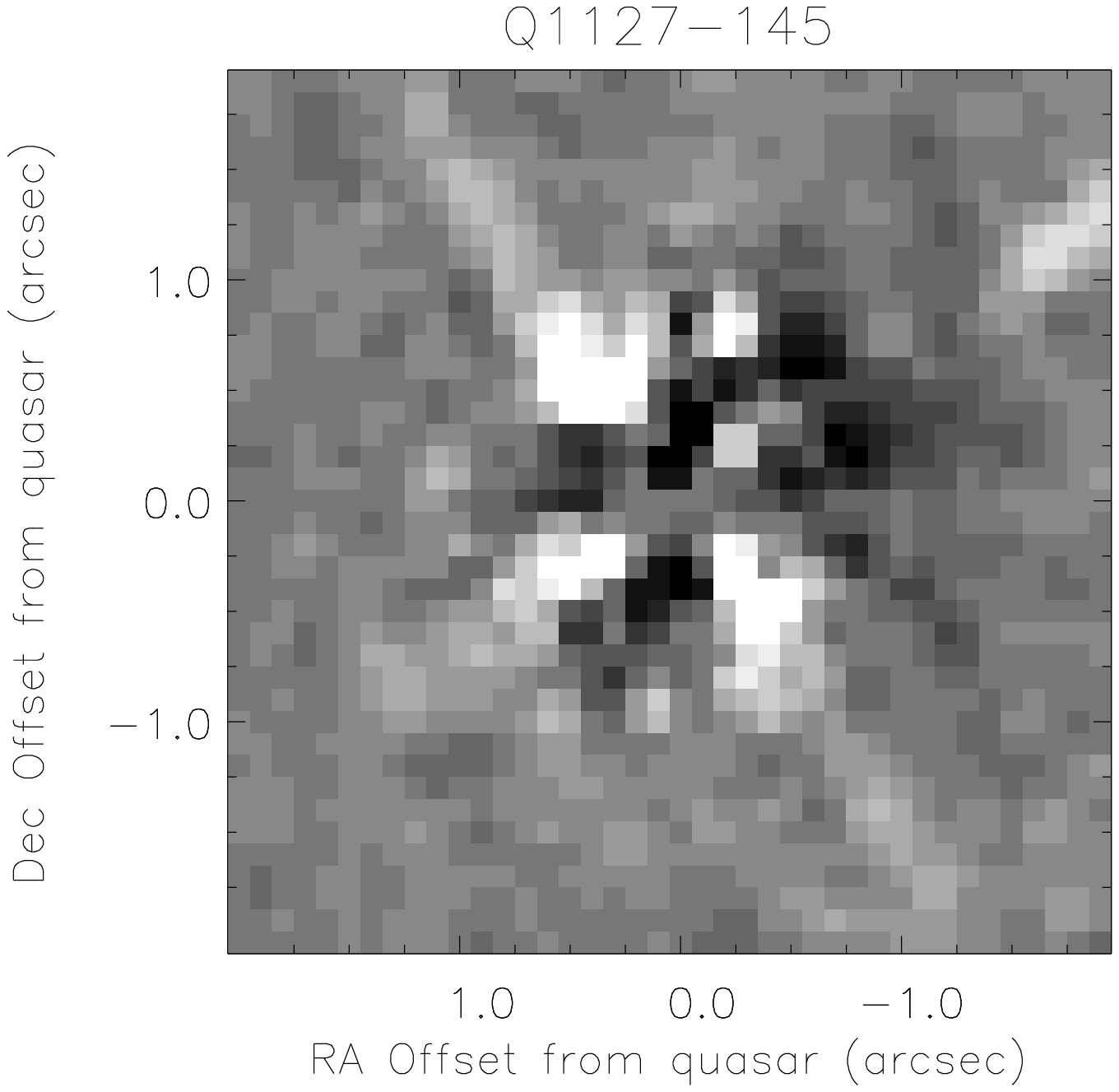}
\end{tabular}
\caption{The upper panels show our image of the field before (left) and
after (right) subtraction  of the QSO image.   The upper-left image 
is $23\arcsec \times 20\arcsec$  corresponding to  $\approx 100 \times 88$ 
kpc$^{2}$ at $z = 0.3127$. Five of the objects can be seen in the full frame 
image.  The sixth object -00.09+00.45, just north of the quasar, is only seen 
after the PSF subtraction.  The emission to the south of the quasar is not
considered since it is less than 0\farcs5 from the quasar.}
\label{fig7}
\end{figure}

\begin{figure}
\begin{tabular}{cc}
\includegraphics[width=0.5\textwidth]{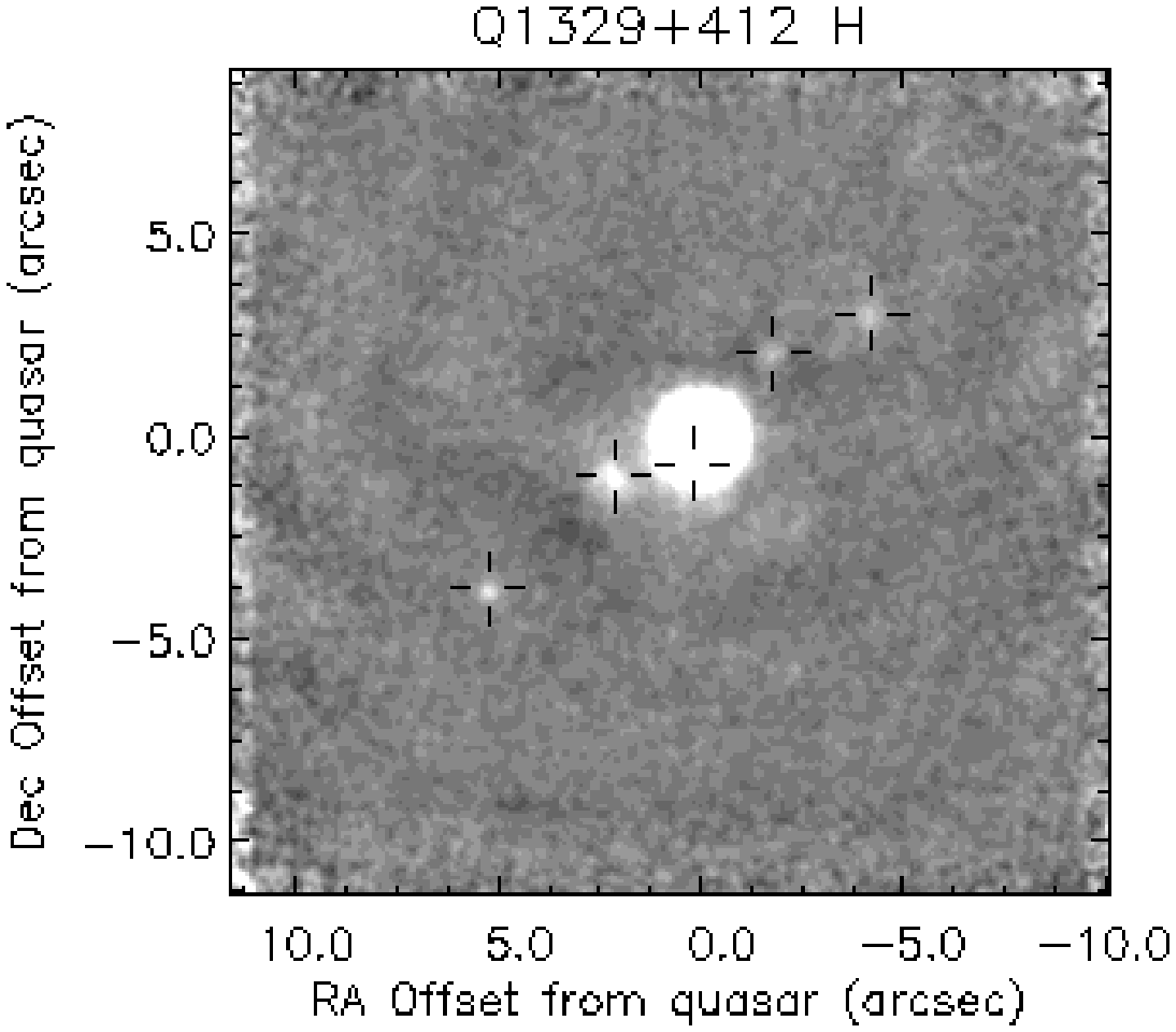}
\includegraphics[width=0.5\textwidth]{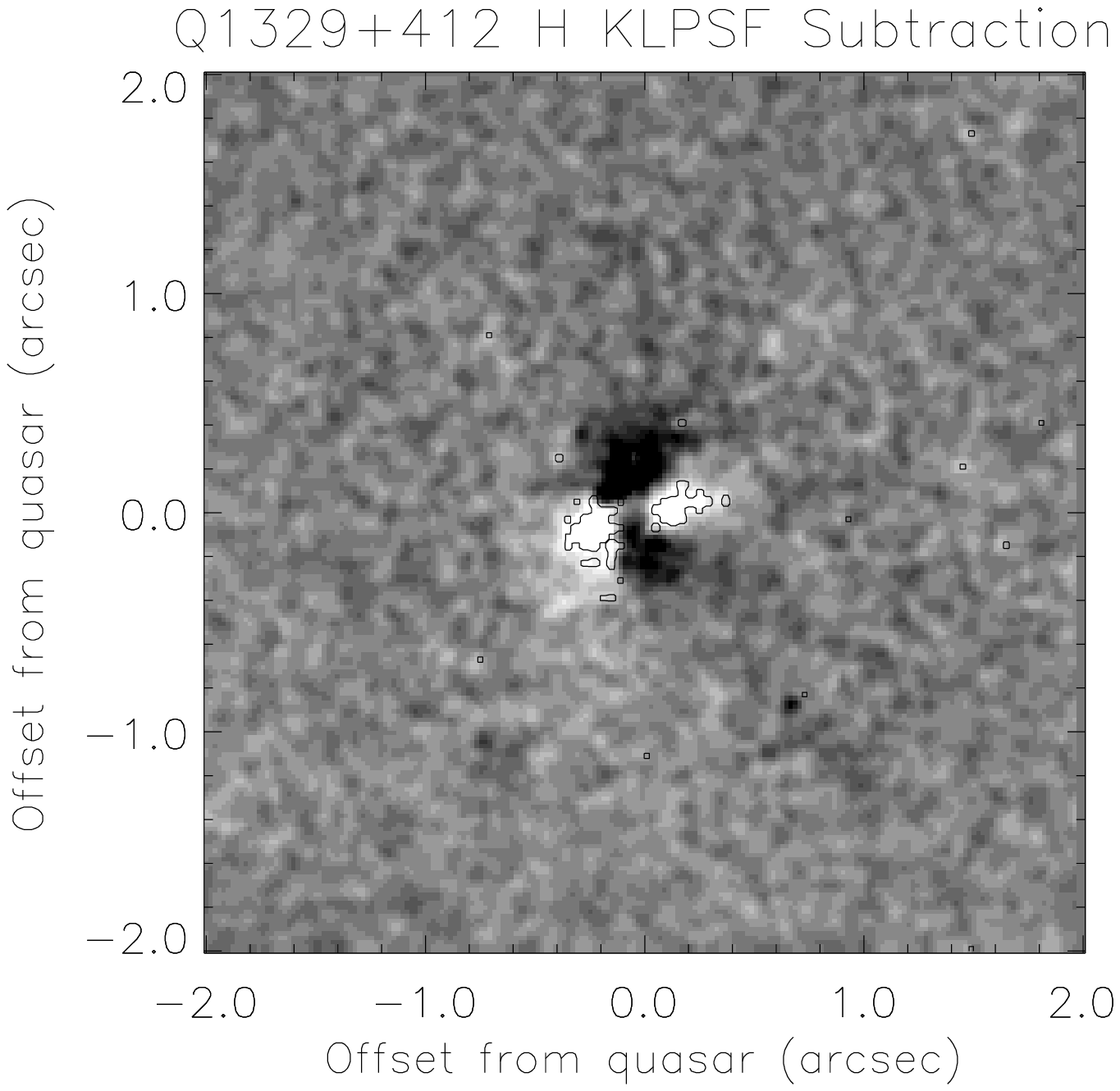}\\
\includegraphics[width=0.5\textwidth]{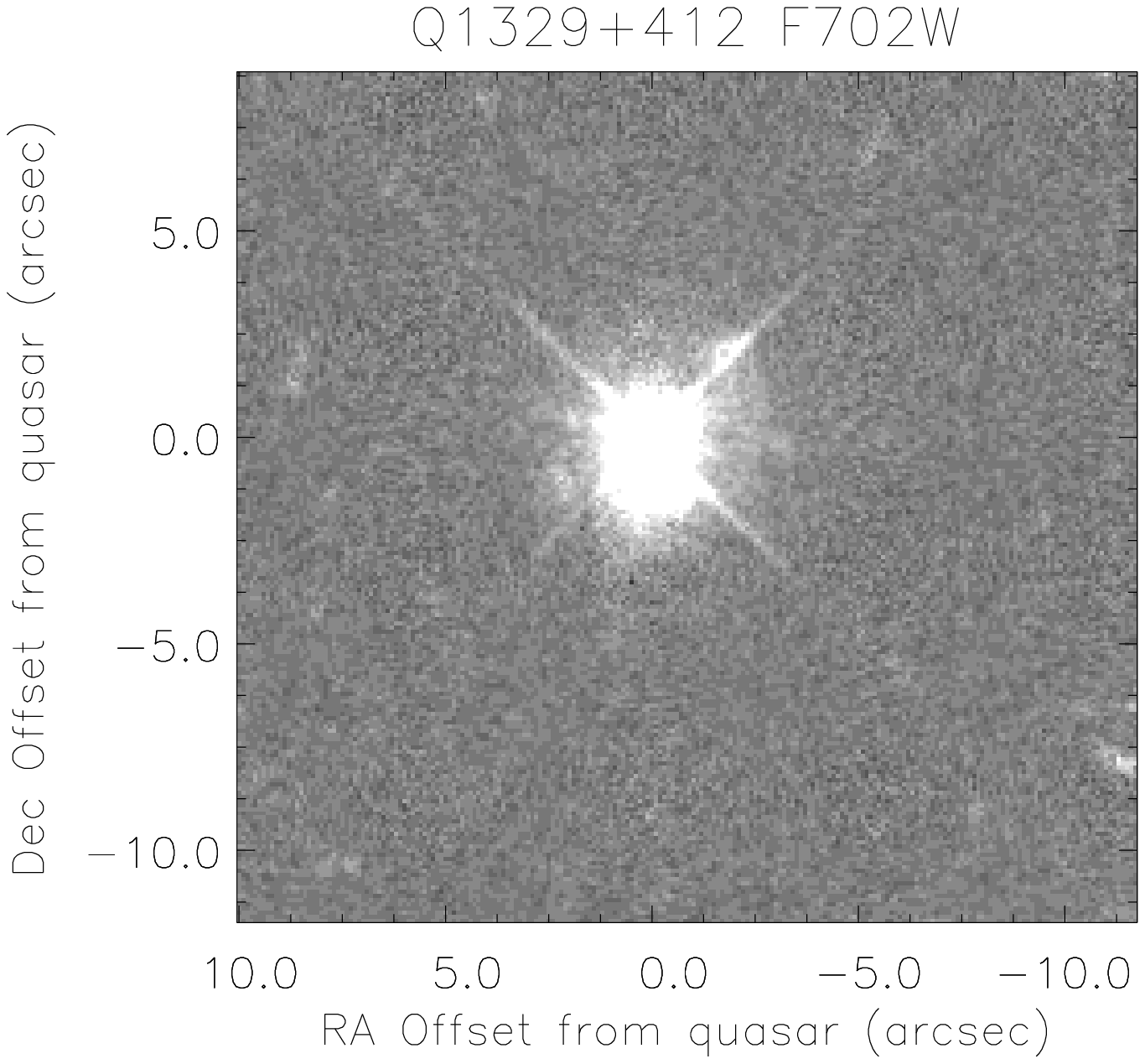}
\includegraphics[width=0.5\textwidth]{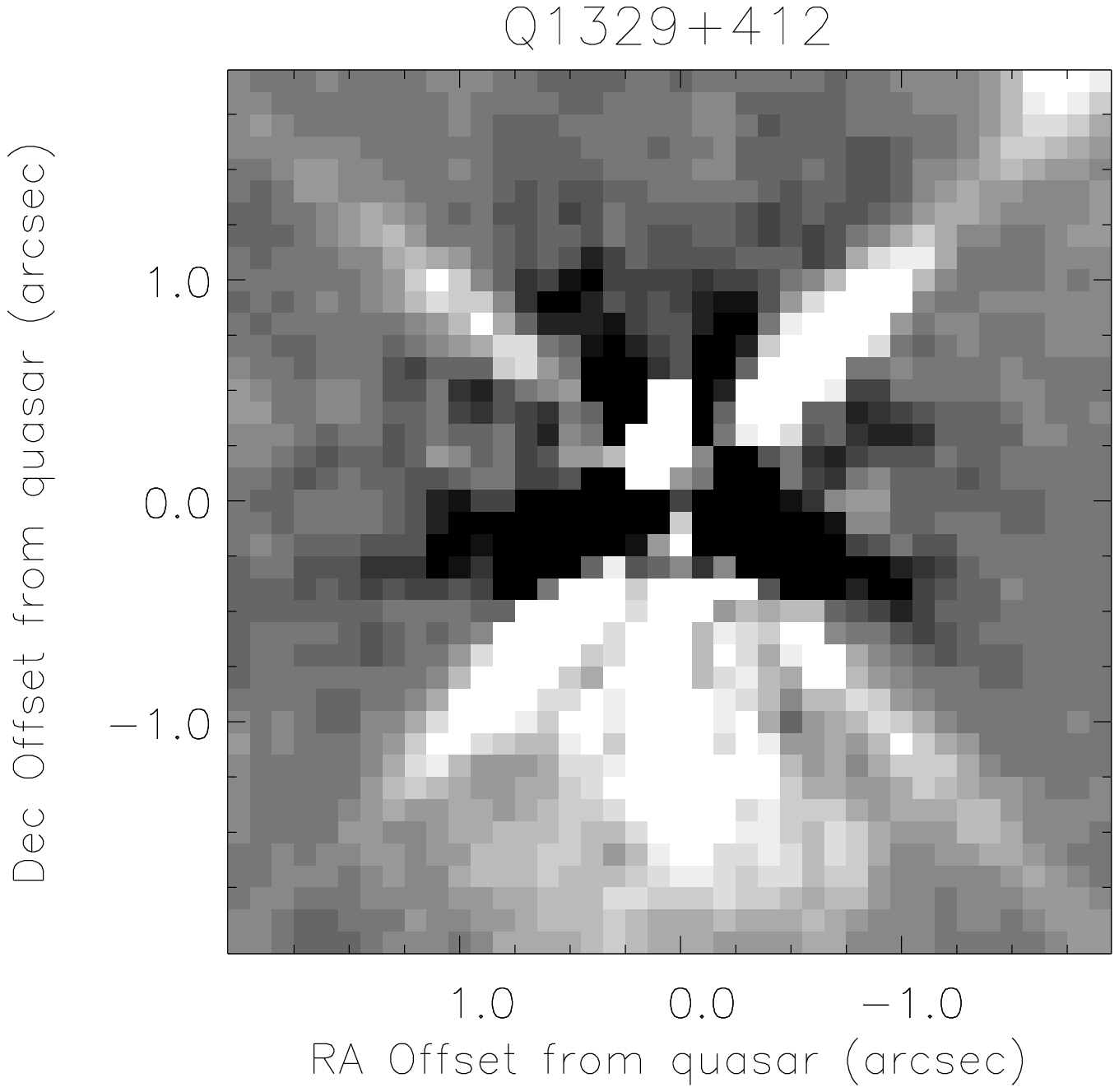}
\end{tabular}
\caption{The upper left and right panels show our image of Q1329+412 before
and after QSO subtraction. The full image (left) is $27.1\arcsec \times
20\arcsec$  corresponding to $\approx 190 \times 165$ kpc$^{2}$ at $z = 1.28$.  
Four objects found by SExtractor are clearly seen in the full frame 
H-band image.   Object -01.79+02.02 is also evident in the HST F702W image 
NW of the quasar (lower right panel).  The H-band and HST QSO-subtracted 
images show similar residual emission.  SExtractor identifies a fifth object 
+0.20-0.69 in the KL QSO subtracted H-band image.  This may correspond to 
the emission seen south of the quasar in the HST image.\label{fig:1329+412}  }
\label{fig8}
\end{figure}

The faint arm and jet-like features discussed by \cite{Turnshek01} are not 
apparent in  our unbinned image.  While our image has an angular resolution of
about $0.2\arcsec$, our $1\sigma$  per pixel limiting surface brightness is
18.9 mag/arcsec$^2$.  The arm and jet like features discussed by
\cite{Turnshek01} are about two magnitudes fainter.   In order to achieve this
sensitivity, we smoothed the image with a gaussian with a FWHM equal to the
FWHM of the quasar image, subtracted an azimuthally-averaged PSF,  and rebinned
the image to $0.2\arcsec$ pixels.  The  resulting image is shown in the
lower-right panel of Figure \ref{fig:0738+313}.  In this image we used an
azimuthally averaged PSF to subtract the light from the quasar because the KL
analysis could not be performed over this  large a field due to its
computational requirements.  The quasar contribution at the separation of the
jet and arm are small so this subtraction should be adequate.   This image was
then analyszed by SExtractor. The feature +5.40-0.11 east of the quasar is
aligned with the bright knot in the \cite{Turnshek01} ``arm'' but the linear
feature WSW of the quasar is found to be slightly rotated from the position
shown in Figure 1 of \cite{Turnshek01}.  In our image the linear feature
extends along a line intersecting the quasar whereas in \cite{Turnshek01} the
feature appears aligned east-west.  The jet clearly shows a highly mottled
linear morphology.

A new feature, +0.71+01.63, can be seen in the smoothed and binned image
(Figure \ref{fig:0738+313}).  It is also seen in the KL PSF subtracted image
when it is  similarly  smoothed and rebinned.   This feature is previously
unidentified.   It appears to have a core with faint emission extending several
arcseconds to the NNW.  It is not clear whether this emission is some component
of a larger object encompassing  all the faint nebulosity but given its close
proximity to the quasar line of sight (b=3 or 6 kpc), it is likely to be
associated with one of the  absorption line systems.   

The feature seen less than $0.5\arcsec$ SW of the quasar line of sight in the
KL PSF-subtracted image is a possible source but it is well within the region 
where PSF subtraction artifacts are large for stellar sources so it is
difficult to rule out that it is a PSF subtraction artifact.

\subsection{Q0850+440}

This radio-quiet QSO has an associated absorption system at $z = 0.1638$.  In
an imaging and spectroscopic survey of faint  galaxies, Lanzetta et al. (1995)
reported a strong Ly-$\alpha$ absorption system and a possible  indication of
weak C IV absorption at $z = 0.1630$. They also obtained an  unambiguous
redshift of $z = 0.1635$ for a galaxy relatively close to  the quasar
sightline. The subsequent  ultraviolet spectroscopic survey of Lanzetta et al.
(1997) showed that  the Ly-$\alpha$ absorption system at $z = 0.1638$  has
$\log N$(H~I) = 19.8. They concluded that  this system is associated with a 
moderate-luminosity early-type galaxy, although it may actually arise in one 
of several very faint galaxies close to the QSO line of sight seen in their
HST/WFPC2 images.   These conclusions are supported by Chen et al. (2001) 
who confirmed the DLA system and the associated galaxy.   Their 
photometric redshifts show that other galaxies in this field do not have 
the same  redshift as the DLA system.

Four objects are identified in our H-band images.  Object -9.04+1.53 is the
galaxy designated as ``G1'' \cite{Lanzetta97}.  We find it is a disk-like galaxy
with a bulge-to-disk ratio of $\sim0.3$ and scale lengths of $r_e\sim0.7$ kpc
and $r_d\sim1.0$ kpc.  Object -0.20-3.49 is the object designated ``S1''
\cite{Lanzetta97}.   It is a point source in our images as well as the HST
images of \cite{Lanzetta97}.   Object +1.28+2.55 is a diffuse arm-like feature
that can be see in our H-band image as well as the HST images.  The NIR
emission appears to extend towards the quasar.  We find an apparent H magnitude
of 24.7 for this diffuse emission while \cite{Lanzetta97} find an apparent AB
magnitude of $m_{F702W}=22.5$ mag.  The F702W-H color of -2.2 is extremely
blue.    Object +0.56+0.32 is close to  the line of sight to the quasar
(0.64\arcsec) but we believe it to be real since we do  not see a similar
extension in the PSF subtracted stellar image (see Figure
\ref{fig:PSFSubtraction}).  In fact we see extended emission to the east of the
quasar in each of the four techniques used to remove the quasar light
contribution.  In addition, the emission at +1.28+2.55 and at +0.56+0.32 
appears to be continuous (Figure 5).  We interpret the two emission features to
be a  single object with object +1.28+2.55 being an extension of the emission. 
We regard it as the likely absorber.  If this is the correct interpretation,
then the DLA is sampling a region very close $\sim$2 kpc from the center of a
very blue galaxy.  In addition, object -2.56+2.01, while identified as a
seperate  object, is also very blue and could be part of this same emission. 
Object -2.56+2.01 is  unresolved.  We do not  detect the object -00023+00043
identified in \cite{Lanzetta97}.

\subsection{Q1127-145}

PKS B1127-145 is a compact, gigahertz-peaked radio source at $z_{em}$ = 1.187
with a jet seen in radio and X-ray images, and variability at radio
wavelengths. Bergeron et al. (1991) identified Mg II, Fe II and Mg I absorption
in the spectrum of the quasar at z = 0.313. They spectroscopically confirmed
two late-type galaxies at the redshift of the absorber separated from the
quasar by 9\farcs6 and 17\farcs7 and identified the closer one as the Mg II
absorber.  Lane et al. (1998) in a survey of H I 21-cm absorption in Mg
II-selected systems using WSRT, discovered 21-cm absorption at z = 0.3127. 
HST/UV spectra show a damped Lyman-$\alpha$ profile with  N(H~I) =  $(5.1 \pm
0.9) \times 10^{21}$ cm$^{-2}$ (Lane et al, 1998; Rao \& Turnshek, 2000). Lane
et al. (1998) concluded that the galaxy which Bergeron et al. (1991) identified
as the absorber, is unlikely to be the DLA system since its column density is 
unlikely to arise at the projected impact parameter ($\ge20$ kpc).   Instead
they suggested that the absorption comes from another galaxy with a separation
3\farcs9 from the quasar, or from tidal debris associated with a group of
galaxies. Bechtold et al (2001) detected X-ray absorption with Chandra/ACIS,
and suggested that the absorbing gas of the DLA has metallicity of 23$\% $
solar.   Rao et al. (2003) identified the DLA galaxy as a patchy/irregular LSB
structure which encompasses four objects.  They suggest that the DLA system
is more likely associated with the faintest object in the group found at the
absorber redshift. Chen \& Lanzetta (2003) also found that a group of at least
four galaxies are at the redshift of the DLA and they concluded that because of
the proximity of these galaxies to the QSO line of sight, it is difficult to
separate the contribution of either of the galaxies to the DLA.

Six objects are identified in our image of this field.  The angular distances
of these  objects to the QSO range from $0\farcs6$ to $16\arcsec$ 
corresponding to b$_{abs}$ = 2.5   to 70 kpc.   Object +8.86+3.98 corresponds
to the object ``G1'' in Bergeron et al. (1991) at $9\farcs7$.  Morphologically
G1 appears to have both a disk and a bulge with a bulge-to-total ratio of
$\sim0.4$.   Tidal warping at the edge of this galaxy can be seen both in our
H-band image and the HST F814W  image.  Our object -3.57+0.17  corresponds to
the object identified as the likely absorber by \cite[]{Lane98}.   

Our image adds to the already  crowded field of Q1127-145.   We regard the
object -0.13+0.57, appearing after PSF subtraction at an angular distance
$0\farcs6$, as another candidate absorber simply due to its close proximity to
the quasar line of sight.  This object has not be indentified previously though
the HST image has a shows a similar feature when the PSF is removed.  We 
do not discount that the faint diffuse emission
seen around the field could also contribute to the absorbing system but this
close object would have an impact parameter of only $\sim$ 2.5 kpc.  It has not
been reported previously  but appears to extend at least one arcsecond away
from the quasar.     We do not detect all of the faint nebulosity  seen in the
immediate vicinity of the quasar in the HST images but have detected very
diffuse emission extending NW $0\farcs5$ from the quasar.   

\subsection{Q1329+412}

This radio-quiet QSO ($z_{em}=1.93$) was observed by Sargent et al. (1988), who
identified four distinct absorption redshifts in its spectrum. In a spectral
survey of C IV absorption systems, they found a weak Mg II doublet at $z =
0.5009$. The IUE spectrum of this object shows a low-redshift candidate DLA
system at $z = 0.5193$ with log $N(H~I) = 20.8$ \cite[]{Lanzetta95}.  HST/UV
spectroscopy did not confirm  the presence of this system (Turnshek et al.
2002) however subsequent HST spectra  did reveal a DLA at the redshift ($z =
1.282$) of another Mg II system (Bechtold et al. 2002).   Based on the
equivalent width of the Ly$\alpha$ line at $z=1.282$, the DLA system has a log
$N(H~I)$ = 19.7. There are additional metal-line systems at $z=1.6012, 1.8355$
(C IV+Mg II), and $z=1.4716, 1.9405$ (C IV).

Five objects are identified in our image of this field.  The angular distances
of these  objects to the QSO range from $\Delta \theta  = 2.4\arcsec$ to
$6.5\arcsec$  corresponding to $b_{abs} = 20$ to $54$ kpc at $z=1.28$. There is
no previous report of  detection of  these objects in the literature.
\cite{Aragon94} detected  a faint object ($K \approx 19.75$) with an angular
distance of $3\arcsec $ from the  QSO line of sight, which could be object
1.81+2.04.    Object -1.81+2.04 is  evident in the HST F702W
image.  All objects are compact and faint.  Objects  -4.23+3.04 and +2.11-0.94
appear slightly extended; however, they would have impact parameters in excess
of 20 kpc if they give rise to the absorption.  We regard the likely absorber
as the faint  emission 0\farcs7 south of the quasar.  This object is evident in
both our NIR image and in the HST image though we do not have the resolution or
the signal-to-noise  to determine its morphology.

\section{SUMMARY AND FUTURE WORK}

We present the first adaptive optics observations of low-redshift DLAs. The
images have revealed several objects at close angular separations  to
the quasar in each field.   The adaptive optics images are comparable to
the HST images in resolution and several close features are seen in common
with HST and with these adaptive optics images.  In addition, we report the 
detection of two previously unidentified objects in the fields of Q0738+313,
where there are no HST images, and Q1127-145, where the HST detection is 
marginal.

The objects found around in these quasar absorption fields would be less than
$0.1L_{*}$ if they are at the absorber redshift and most of the brighter
objects appear to have disks. The census of the brighter objects in these six
absorber fields is 6 disk-dominated galaxies, 3 disk+bulge galaxies,  2
bulge-dominated galaxies, 3 point sources, and 5 unconstrained objects.  In
addition, five of the six fields show objects between 0\farcs5 and 1\farcs0 to
the line of sight to the quasar.  

Our census has found likely candidates for all of the DLA systems.  The KL 
subtraction reveals a candidate object just offset from the quasar line of
sight in Q0054+144 though the HST field appears to have several faint objects 
distributed about the field.  The DLA in Q0235+164 appears to be the object
previously identified 1\arcsec East of the quasar \cite[]{Yanny89}.  In
Q0738+313 we find a new object to which we attribute the lower-redshift DLA.  
This object would have an impact parameter of $\sim$3 kpc.  It appears to have
emission extending several arcseconds to the NW.   This emission could be
associated  with the jet and arm features identified by \cite{Turnshek01}
though this  emission is fainter than the new object detected here.  The DLA at
$z=0.22$ in this field has been previously identified.  In Q0850+440 
\cite{Lanzetta95} find a dwarf galaxy 9\arcsec~ from the quasar ($b\sim25$
kpc).  We identify another object very close to the quasar line of sight as a
candidate absorber.  It appears only after subtraction of the quasar but if the
absorption arises from this object, then the DLA arises close to the core of a
very blue galaxy ($b\sim2$ kpc).  For Q1127-145 we find a faint diffuse  object
close to the line of sight of the quasar and extending NNW several arcseconds. 
The absorber in Q1329+412 is identified as arising from  an object 0\farcs7
south of the quasar.  This object is also clearly seen in both our H-band image
and an HST F702W image of the field.

All candidate absorbers are faint, with luminosities less than 0.1 $L_{H}^{*}$
or $L_{K'}^{*}$.  Assuming that at least some of these objects are at the same
redshift as the absorbers, we conclude that the  absorbers in our fields are
associated with relatively low luminosity galaxies. Morphological analysis
reveals that most of the brighter objects have a disk component.  Their sizes,
inferred from the surface brightness profiles, range from small to typical
scale lengths for local disk galaxies.   For reference, our Galaxy has a disk
scale length of $r_d = 2.2$~kpc measured in the $K$-band
\cite[]{maihara78,jones81} and M31 has a scale length of $r_d = 3.9$~kpc in the
$K$-band, and $r_d = 4$~kpc in the $I$-band \cite[]{hiromoto83}.  Table 5
summarizes object morphologies and the  derived linear impact parameter,
luminosity, and scale lengths  assuming the objects are at the redshift of the
absorber.  

\cite{Rao03} have suggested  that the DLAs at $z < 1$ are dominated by
dwarf or low surface brightness  galaxies. However, \cite{Chen03},
with more photometric redshifts,  have suggested that the luminosity function
of $z < 1$ DLAs could be much broader.  Our observations, at higher resolution
than both of these studies, have  found all of the candidate absorbers to be
faint, with significant  disk components for the majority of the objects. This
suggests that  a considerable fraction of low-$z$ DLAs may be faint, low
surface brightness galaxies.  Such a conclusion would appear to be consistent
with the low metallicities found  in low-$z$ DLAs (e.g., \cite{Khare04};
\cite{Kulkarni05}; and references therein). However, it would be necessary
to obtain redshift confirmations for our  candidates and to obtain similar
high-resolution images of other low-$z$ DLAs to reach more definitive
conclusions on the luminosity function of the absorber galaxies. 

Our observations  have demonstrated the use of adaptive optics for direct
high-resolution imaging  of the galaxies giving rise to quasar absorbers. 
Deeper observations of the same fields in the future with higher order  AO
systems would help to improve the signal-to-noise ratios  in the fainter
objects. Furthermore, adaptive  optics systems with laser guide stars
are not constrained by the  need to have a bright guide star in the quasar
field, and would thus be able to reach  higher redshift absorbers. 

It is crucial to also obtain spectroscopy (or at least narrow-band imaging)  of
all the fields to better constrain the redshifts of the  detected candidate
absorbers. With spectroscopic PSF subtraction  procedures (such as those
followed by \cite{Moller00a}) it may be  feasible to even verify the redshifts
of the objects located very close to the  line of sight of the quasar.  It is
essential to expand the sample of high-resolution broad-band images,  followed
with spectroscopic confirmations, for quasar absorbers at low and  high
redshifts. Such a combination of high-resolution imaging and spectroscopic 
observations of quasar absorbers can give direct information  on their
luminosities, sizes,  and star formation rates and thus the nature of these
galaxies.  Performing such observations on different types of  quasar absorbers
(e.g., DLAs, weak Mg II systems, C IV systems) may help to understand any
trends between the absorption line  strengths and galaxy properties such as the
luminosities and impact parameters.  Finally, a comparison of these properties
of quasar absorbers  at low and high redshifts will allow us to study the
evolution of the  absorbing galaxies with cosmological time and 
the connection between the absorbers and the present-day galaxies.

\acknowledgments

This paper is based on observations obtained at the Gemini Observatory, which
is operated by the Association of Universities for Research in Astronomy, Inc.,
under a cooperative agreement with the NSF on behalf of the Gemini partnership:
the National Science Foundation (United States), the Particle Physics and
Astronomy Research Council (United Kingdom), the National Research Council
(Canada), CONICYT (Chile), the Australian Research Council (Australia), CNPq
(Brazil) and CONICET (Argentina).  This paper is based on observations obtained
with the Adaptive  Optics System Hokupa'a/Quirc, developed and operated by the 
University of Hawaii Adaptive Optics Group, with support from the  National
Science Foundation.  We thank the Gemini-North  Observatory staff for
assistance during our observations,  Hsiao-Wen Chen for providing details of
the published photometric data for Q1127-145, and B. Stobie of the University
of Arizona for providing and assisting with  the IDP-3 package.  We also thank
the referee of the paper for making a number of positive suggestions on the
paper.  VPK and SG gratefully acknowledge partial support from the  National
Science Foundation grant AST-0206197 and from the  University of South Carolina
Research Foundation.   MT acknowledges support from National Science Foundation
grant AST-0205960.



Facilities: \facility{Gemini-N(Hokupa'a)}.




\clearpage



 \clearpage

\clearpage

\clearpage

\clearpage


\clearpage


\clearpage

\clearpage







\clearpage
\begin{deluxetable}{lllllll}
\tablecaption{Target Properties\label{tab:prop}}
\tablewidth{0pt}
\tablehead{
\colhead{QSO} & 
     \colhead{V} & 
     \colhead{$z_{em}$} & 
     \colhead{$z_{abs}$} & 
     \colhead{$\log N$(H~I)} &
     \colhead{Absorber type} & 
     \colhead{N(H~I) Reference\tablenotemark{a}} 
}
\startdata
Q0054+144 &16.1      &0.171    &0.103    &$18.3$	& Lyman-limit	&1, 2\\
Q0235+164 &15.5-19   &0.940    &0.524    &$21.65$	& DLA    	&3\\
Q0738+313 &16.1      &0.635    &0.0912   &$21.18$	& DLA    	&4, 5\\
Q0738+313 &16.1      &0.635    &0.2212   &$20.90$	& DLA    	&4, 6\\
Q0850+440 &16.4      &0.5139   &0.1638   &$19.81$	& sub-DLA	&7\\
Q1127-145 &16.9      &1.184    &0.3127   &$21.7$	& DLA     &3\\
Q1329+412 &16.3      &1.9300   &1.282    &$19.7$	& sub-DLA	&8\\
\enddata

\tablenotetext{a}{References: 1. Lanzetta et al. (1995); 2. Turnshek et al. 
(2002); 3. Junkkarinen et al. (2004); 4. Rao \& Turnshek (1998);  5. Chengalur 
\& Kanekar (1999); 6. Lane et al. (1998); 7. Lanzetta et al. (1997).; 8.  Bechtold et al. (2002);
}

\end{deluxetable}

\begin{deluxetable}{ccccccc}
\tablecaption{Summary of Observations\label{tab:obs}}
\tablewidth{0pt}
\tablehead{
\colhead{QSO} & 
     \colhead{Filter} & 
     \colhead{Total Integration Time} &
     \colhead{FWHM\tablenotemark{a}} & 
     \colhead{50\%-EED\tablenotemark{a}} & 
     \colhead{$1\sigma~ \mu_{lim}$\tablenotemark{b}} & 
     \colhead{$3\sigma~ m_{lim}$\tablenotemark{c}} \\
\colhead{} & 
     \colhead{} & 
     \colhead{seconds} &
     \colhead{arcsec} & 
     \colhead{arcsec} & 
     \colhead{mag/arcsec$^2$} & 
     \colhead{mag}
}
\startdata
Q0054+144 & H    & $23\times120$ & 0.46 & 0.75 & 19.6 & 21.9 \\
Q0235+164 & H    & $53\times120$ & 0.50 & 0.82 & 19.9 & 22.2 \\
Q0738+313 & K$'$ & $74\times60$  & 0.19 & 0.28 & 18.9 & 21.2 \\
Q0850+440 & H    & $39\times120$ & 0.23 & 0.38 & 20.0 & 22.3 \\
Q1127-145 & H    & $32\times180$ & 0.33 & 0.54 & 20.0 & 22.3 \\
Q1329+412 & H    & $48\times180$ & 0.29 & 0.47 & 20.1 & 22.4 \\
\enddata  

\tablenotetext{a}{FWHM and 50\% EED measured from the QSO.}
\tablenotetext{b}{$1\sigma$ per pixel limit}
\tablenotetext{c}{$3\sigma$ limit within  75 pixels }
\end{deluxetable}

\begin{deluxetable}{llll}
\tablecaption{Summary of Objects Detected\label{tab:obj}}
\tablewidth{0pt}
\tablehead{
\colhead{QSO} 
     & \colhead{Object$\Delta \alpha (\arcsec)$$\Delta \delta (\arcsec)$} 
     & \colhead{$\Delta \theta (\arcsec)$}
     & \colhead{m(AB)} 
}
\startdata
Q0054+145
               & QSO          &    0    & $H=14.6$\\
               & -0.29-0.74   &   0.79  & 21.5\\
&&&\\
Q0235+164
               & QSO          &    0    & $H=16.7$\tablenotemark{\dagger} \\
               & -0.33-0.45   &   0.56  & 23.9\tablenotemark{\dagger} \\
               & +1.11-0.01   &   1.11  & 21.9\tablenotemark{\dagger} \\
               & +0.15-1.91   &   1.92  & 20.2\tablenotemark{\dagger} \\
               & +2.40+0.93   &   2.57  & 24.9\tablenotemark{\dagger} \\
               & -5.85-2.69   &   6.44  & 20.3\tablenotemark{\dagger} \\
               & -4.93-7.47   &   8.95  & 24.3\tablenotemark{\dagger} \\
               & -7.15-7.21   &  10.15  & 20.6\tablenotemark{\dagger} \\
&&&\\
Q0738+313      & QSO          &    0    & $K'=16.1$ \\
               & +0.71+1.63   &   1.78  & 22.6\tablenotemark{a}\\
               & +2.02+1.54   &   2.54  & 21.3\\
               & +5.40-0.11   &   5.40  & 23.0\tablenotemark{a}\\
               & +1.90-5.38   &   5.71  & 19.8\\
               & -6.58-1.64   &   6.78  & 22.4\tablenotemark{a,b}\\
               & -7.96+5.02   &   9.41  & 21.4\\
 
&&&\\
Q0850+440      & QSO          &    0    & $H=15.8$ \\
               & +0.56+0.32   &   0.64  & ...\\
               & +1.28+2.55   &   2.85  & 24.7\tablenotemark{c}\\
               & -2.56+2.01   &   3.25  & 25.8\tablenotemark{c}\\
               & -0.20-3.49	&   3.50  & 17.6 \\
               & -9.04+1.53	&   9.17  & 17.8 \\
%
%
&&&\\
Q1127-145      & QSO          &    0    & $H=16.5$ \\
               & -0.13+0.57   &   0.58  & ... \\
               & -3.57+0.17	&   3.57  & 24.6\tablenotemark{d} \\
               & +5.42-1.40	&   5.60  & 22.6 \\
               & +7.92+1.82	&   8.13  & 23.8 \\
               & +8.86+3.98   &   9.71  & 18.5\\              
               & +14.49-6.76  &  16.0   & 19.9\\
&&&\\
Q1329+412      & QSO          &    0    & $H=17.0$ \\
               & +0.20-0.69   &   0.72  & ...\\
               & +2.11-0.94	&   2.31  & 22.5 \\
               & -1.81+2.04   &   2.73  & 25.7\tablenotemark{c}\\
               & -4.23+3.04	&   5.21  & 24.6\tablenotemark{e}\\
               & +5.23-3.76	&   6.44  & 24.9 \\
%
%
%

\enddata
\tablenotetext{\dagger}{Observations were nonphotometric for 0235+164. 
Magnitudes are given only for relative photometry.}
\tablenotetext{a}{These objects are only detected after the image was heavily
smoothed by a gaussian FWHM=0\farcs2.  The detections are $\ge 3\sigma$/pixel
in the smoothed image.  Object +5.40-0.11 corresponds to the \cite{Turnshek01}
``arm'' while object -6.58-1.64 corresponds to their ``jet''.}
\tablenotetext{b}{SExtractor also found an object at -1.38-0.76.  This object
has a magnitude of 24.7.  We have combined it the linear feature listed
here.}
\tablenotetext{c}{Detected with 75 pixels, $0.9\sigma$/pixel threshold}
\tablenotetext{d}{Combined three close detections}
\tablenotetext{e}{Detected with 75 pixels, $0.8\sigma$/pixel threshold}
\end{deluxetable}

\clearpage

\begin{deluxetable}{llllllll} 
\tablecaption{Morphological Parameters\label{tab:morph}}
\tablewidth{0pt}
\tablehead{
     \colhead{QSO} &
     \colhead{Obj} &
     \colhead{$B/T$} &
     \colhead{$r_{1/n}$} &
     \colhead{$r_{d}$} &
     \colhead{$n$} \\

     \colhead{}    &
     \colhead{}    &
     \colhead{}      &
     \colhead{$(\arcsec)$} &
     \colhead{$(\arcsec)$} &
     \colhead{} \\ 
}
\startdata
 Q0235+164 &&&&&&&\\
  &  $+1.11-0.01$  & $0.50_{-0.50}^{+0.50}$ & $0.00_{-0.00}^{+0.01}$ & $0.05_{-0.05}^{+0.01}$ & $3.29_{-1.88}^{+2.09}$ \\
  &  $+0.15-1.91$  & $0.27_{-0.03}^{+0.03}$ & $0.00_{-0.00}^{+0.00}$ & $0.21_{-0.01}^{+0.00}$ & $5.02_{-0.31}^{+0.30}$ \\
  &  $-5.85-2.69$  & $0.26_{-0.08}^{+0.08}$ & $0.02_{-0.02}^{+0.02}$ & $0.20_{-0.02}^{+0.02}$ & $5.21_{-0.35}^{+0.55}$ \\
  &  $-4.93-7.47$  & $0.20_{-0.20}^{+0.58}$ & $0.09_{-0.09}^{+0.18}$ & $0.10_{-0.07}^{+0.18}$ & $3.49_{-0.20}^{+0.40}$ \\
  &  $-7.15-7.21$  & $0.91_{-0.14}^{+0.09}$ & $0.33_{-0.08}^{+0.09}$ & $0.33_{-0.32}^{+0.09}$ & $3.81_{-0.90}^{+0.67}$ \\
 Q0738+313 &&&&&&&\\
  &  $+2.02+1.54$  & $1.00_{-0.70}^{+0.00}$ & $0.00_{-0.00}^{+0.00}$ & $0.01_{-0.00}^{+0.00}$ & $7.32_{-2.47}^{+2.68}$ \\
  &  $+1.90-5.38$  & $0.34_{-0.11}^{+0.11}$ & $0.03_{-0.02}^{+0.02}$ & $0.15_{-0.02}^{+0.02}$ & $3.32_{-0.20}^{+0.24}$ \\
  &  $-7.96+5.02$  & $0.68_{-0.68}^{+0.32}$ & $0.00_{-0.00}^{+0.01}$ & $0.00_{-0.00}^{+0.01}$ & $7.38_{-2.93}^{+2.62}$ \\
 Q0850+440 &&&&&&&\\
  &  $-2.56+2.01$  & $0.11_{-0.11}^{+0.21}$ & $0.01_{-0.01}^{+0.08}$ & $0.22_{-0.10}^{+0.08}$ & $3.70_{-0.79}^{+1.30}$ \\
  &  $-0.20-3.49$  & $0.00_{-0.00}^{+0.00}$ & $0.10_{-0.00}^{+0.00}$ & $0.00_{-0.00}^{+0.00}$ & $3.94_{-0.01}^{+0.00}$ \\
  &  $-9.04+1.53$  & $0.24_{-0.01}^{+0.01}$ & $0.24_{-0.01}^{+0.01}$ & $0.40_{-0.00}^{+0.01}$ & $3.69_{-0.08}^{+0.09}$ \\
 Q1127-145 &&&&&&&\\
  &  $+5.42-1.40$  & $0.96_{-0.39}^{+0.04}$ & $0.11_{-0.06}^{+0.18}$ & $0.00_{-0.00}^{+0.18}$ & $3.93_{-1.13}^{+1.25}$ \\
  &  $+7.92+1.82$  & $0.57_{-0.57}^{+0.43}$ & $0.12_{-0.12}^{+0.12}$ & $0.32_{-0.22}^{+0.12}$ & $4.25_{-1.44}^{+1.29}$ \\
  &  $+8.86+3.98$  & $0.36_{-0.03}^{+0.02}$ & $0.22_{-0.04}^{+0.01}$ & $0.88_{-0.07}^{+0.01}$ & $4.09_{-0.21}^{+0.11}$ \\
  &  $+14.5-6.76$  & $0.19_{-0.07}^{+0.10}$ & $0.02_{-0.02}^{+0.10}$ & $0.22_{-0.02}^{+0.10}$ & $3.99_{-0.49}^{+0.36}$ \\
 Q1329+412 &&&&&&&\\
  &  $+2.11-0.94$  & $0.16_{-0.09}^{+0.11}$ & $1.76_{-0.81}^{+1.15}$ & $3.52_{-0.02}^{+1.15}$ & $4.91_{-0.10}^{+0.20}$ \\
  &  $-1.81+2.04$  & $0.56_{-0.56}^{+0.44}$ & $0.00_{-0.00}^{+0.00}$ & $0.27_{-0.16}^{+0.00}$ & $2.41_{-1.24}^{+1.84}$ \\
  &  $-4.23+3.04$  & $0.87_{-0.75}^{+0.13}$ & $0.06_{-0.06}^{+0.15}$ & $0.30_{-0.15}^{+0.15}$ & $5.78_{-0.85}^{+0.62}$ \\
  &  $+5.23-3.76$  & $0.64_{-0.64}^{+0.36}$ & $0.01_{-0.01}^{+0.03}$ & $0.16_{-0.16}^{+0.03}$ & $4.64_{-1.04}^{+1.12}$ \\
\enddata
\end{deluxetable}

\begin{deluxetable}{lllllllll}
\tablecaption{Summary of Derived Quantities of Objects Detected\label{tab:summary}}
\tablewidth{0pt}
\tablehead{
\colhead{QSO} & 
     \colhead{Obj}  & 
     \colhead{b$_{\rm abs}$\tablenotemark{a}} &
     \colhead{$log(L/L_{*})$\tablenotemark{b}} &
     \colhead{Morphology\tablenotemark{c}} & 
     \colhead{$r_{1/n}$\tablenotemark{d}} & 
     \colhead{$r_{d}$\tablenotemark{d}} \\
\colhead{} &
     \colhead{}  & 
     \colhead{[kpc]} &
     \colhead{} &
     \colhead{} & 
     \colhead{$[kpc]$} & 
     \colhead{$[kpc]$} \\
}
\startdata
Q0054+145      & -0.29-0.74 & 1.40         &              &        &           &     \\ 

&&&&&\\
Q0235+164\tablenotemark{e}
               & -0.33-0.45 & 3.36         &              &        &           &     \\ 
               & +1.11-0.01 & 6.66         &              &        & 0.0       & 0.3\\
               & +0.15-1.91 & 11.53        &              & B+D    & 0.0       & 1.3\\
               & +2.40+0.93 & 15.43        &              &        &           &     \\ 
               & -5.85-2.69 & 38.66        &              & Dd     & 0.1       & 1.2\\
               & -4.93-7.47 & 53.73        &              &        & 0.5       & 0.6\\
               & -7.15-7.21 & 60.93        &              & B      & 2.0       & 2.0\\
&&&&&\\
Q0738+313\tablenotemark{f}
               & +0.71+1.63 & 2.90   6.09  & -2.32  -1.48 &        &           &     \\ 
               & +2.02+1.54 & 4.14   8.69  & -1.80  -0.96 & P      & 0.0  0.0  & 0.0  0.0\\
               & +5.40-0.11 & 8.80   18.48 & -2.48  -1.64 &        &           &     \\ 
               & +1.90-5.38 & 9.31   19.54 & -1.20  -0.36 & D+B    & 0.1  0.1  & 0.2  0.5\\
               & -6.58-1.64 & 11.05  23.21 & -2.24  -1.40 &        &           &     \\ 
               & -7.96+5.02 & 15.34  32.21 & -1.84  -1.00 & P      & 0.0  0.0  & 0.0  0.0\\
&&&&&\\
Q0850+440      & +0.56+0.32 & 1.77         &              &        &           &     \\
               & +1.28+2.55 & 7.69         & -4.09        &        &           &     \\
               & -2.56+2.01 & 8.77         & -4.53        & D      & 0.0       & 0.6 \\
               & -0.20-3.49 & 9.44         & -1.25        & P      & 0.3       & 0.0 \\
               & -9.04+1.53 & 24.74        & -1.33        & D      & 0.7       & 1.1 \\
&&&&&\\
Q1127-145      & -0.13+0.57 & 2.55         &              &        &           &      \\
               & -3.57+0.17 & 15.70        & -3.42        &        &           &      \\
               & +5.42-1.40 & 24.63        & -2.62        &        & 0.5       & 0.0 \\
               & +7.92+1.82 & 35.75        & -3.10        &        & 0.5       & 1.4 \\
               & +8.86+3.98 & 42.70        & -0.98        &        & 1.0       & 3.9 \\
               & +14.49-6.76& 70.32        &              &        & 0.1       & 1.0 \\
&&&&&\\
Q1329+412\tablenotemark{g}
               & +0.20-0.69 & 4.30         &              &        &           &      \\ 
               & +2.11-0.94 & 13.80        & -2.06        & D      & 10.       & 21.\\
               & -1.81+2.04 & 16.31        & -3.34        &        & 0.0       & 1.6 \\
               & -4.23+3.04 & 31.13        & -2.90        &        & 0.4       & 1.8 \\
               & +5.23-3.76 & 38.48        & -3.02        &        & 0.1       & 1.0 \\
\enddata
\tablenotetext{a}{Projected impact parameter in units of $h^{-1}$ kpc, assuming the object to be 
at the absorber redshift with $h =0.73$. In the case of Q0738+313, the two numbers given 
correspond to the two absorbers at $z=0.0912$ and $z=0.2212$, respectively.}
\tablenotetext{b}{The object luminosities assume the objects are at the absorber redshifts
and are based on our photometry.  Q0738+313 lists luminosities in K$'$.}
\tablenotetext{c}{Morphology codes:Dd -- Disk dominated; D -- Disk; B+D -- Bulge plus Disk; B -- bulge; P -- unresolved; blank -- unknown type}
\tablenotetext{d}{The profile linear scale length assuming object is at absorber redshift }
\tablenotetext{e}{ Observations were non-photometric. }
\tablenotetext{f}{ Values correspond to $z_{abs} = 0.0912$ and  $z_{abs} = 0.2212$ respectively. }
\tablenotetext{g}{ Values corresponds to $z_{abs}=1.282$ }
\end{deluxetable}
\clearpage

\end{document}